\documentclass[preprint]{elsarticle}
\usepackage{amsmath}
\allowdisplaybreaks[4]
\usepackage{lipsum,soul}
\usepackage{bm}
\usepackage{graphicx}
\usepackage{graphics}
\usepackage{float}
\usepackage{amsmath}
\usepackage{array}
\usepackage{xcolor}
\usepackage{subfig}
\usepackage{hyperref}
\usepackage[capitalise]{cleveref}
\usepackage{textcomp}

\usepackage{epstopdf}
\usepackage{cases}
\usepackage{amssymb}
\usepackage{multirow}
\usepackage{siunitx}
\usepackage{cases}
\usepackage{tabularx}
\usepackage{nomencl}
\usepackage[top=1.1in, bottom=1.1in, left=1.0in, right=1.0in]{geometry}
\usepackage{tikz}
\usepackage{comment}
\usepackage{lineno,hyperref}

\modulolinenumbers[1]
\usepackage{booktabs}

\begin{document}
\begin{frontmatter}

\title{An efficient discrete unified gas kinetic scheme for strongly inhomogeneous fluids at the nanoscale}
\author[University1]{Huipeng Liu}
\author[University2]{Zhaoli Guo\corref{corresponding}} 
\ead{zlguo@hust.edu.cn}

\affiliation[University1]{organization={State Key Laboratory of Coal Combustion, School of Energy and Power Engineering, Huazhong University of Science and Technology},
			city={Wuhan},
			postcode={430074}, 
			country={China}}
\affiliation[University2]{organization={Institute of Interdisciplinary Research for Mathematics and Applied Science, Huazhong University of Science and Technology},
			city={Wuhan},
			postcode={430074}, 
			country={China}}
            
\cortext[corresponding]
{Corresponding author.}
\date{\today}

\begin{abstract}
We develop an efficient discrete unified gas kinetic scheme (DUGKS) to solve a kinetic model for strongly inhomogeneous fluid flows at the nanoscale. The proposed DUGKS adopts efficient numerical strategies to evaluate the multiple nonlocal integrals involved in the kinetic equation, reducing the computational cost from \(O(N N_{\sigma})\) for the direct evaluation to \(O(N)\), where \(N\) and \(N_\sigma\) are the number of cells in the flow domain and integral domain, respectively. The accuracy and efficiency of the proposed DUGKS are assessed through several test cases, including static fluid structures and force-driven flow in nano slits. The present results agree well with those from Monte Carlo and molecular dynamics simulations, as well as with those from the original DUGKS with direct evaluation of the integrals. Moreover, the speedup relative to the original DUGKS can reach up to two orders of magnitude. As example applications, we investigate two-dimensional pressure-driven flow between parallel plates and force-driven flow in a square duct, revealing distinctive nanoscale features such as the non-equivalence of pressure- and force-driven flows and the formation of density peaks in the square duct corners.
\end{abstract}



\end{frontmatter}

\section{Introduction}\label{sec:1}
Fluid flows at the nanoscale appear in a variety of engineering applications, such as carbon sequestration \cite{bourg2015nanoscale}, water purification \cite{shannon2008science}, and shale gas production \cite{shan2021contribution,wang2016molecular}. At the nanoscale, molecular volume and molecular interactions are significant \cite{karniadakis2006microflows}, resulting in strong inhomogeneity in fluid properties. Consequently, the conventional Navier-Stokes equations are not applicable to such flows. Although molecular dynamics (MD) and Monte Carlo methods have been widely employed to study nanoscale fluid systems, they usually suffer from high computational costs, limited temporal and spatial scales, and substantial statistical noise \cite{li2010molecular,khademi2011molecular}. Kinetic theory has gained considerable interest for its capability to accurately describe fluid properties at the nanoscale, while maintaining a significantly lower computational cost \cite{guo2005simple,su2023kinetic,shan2023molecular,shan2025molecular}.

A widely used kinetic model for bulk fluids taking into account of molecular volume and inter-molecular interactions is the Enskog-Vlasov (EV) equation \cite{grmela1971kinetic,karkheck1981kinetic}. However, for nano-confined fluid systems, the EV model becomes invalid due to the strong inhomogeneity arising from fluid-wall and fluid-fluid interactions \cite{vanderlick1989molecular}. Inspired by the density functional theory \cite{nordholm1980generalized,tarazona1984density,tarazona1985free}, Davis et al. \cite{davis1987kinetic,ted1987kinetic} introduced a weighted average density to replace the local density in the EV equation, successfully modeling the density distribution of confined fluids at the nanoscale. Nevertheless, the complicated collision operator in their model limits its practical application. To address this limitation, Guo et al. \cite{guo2005simple,guo2006generalized,guo2005temperature} simplified the collision operator in the EV equation and introduced a nonlocal gradient in place of the local gradient, effectively predicting the density distribution and flow characteristics of confined fluids at the nanoscale.

Both stochastic and deterministic approaches have been devised to solve nanoscale kinetic equations. The direct simulation Monte Carlo method, as a stochastic method, can accurately predict fluid properties but is prone to significant statistical noise due to its stochastic nature \cite{sadr2019treatment,jenny2010solution,nedea2006density}. The deterministic methods, including the DUGKS \cite{shan2020discrete}, the fast spectral method \cite{wu2015fast,wu2016non}, and the discrete velocity method \cite{su2023kinetic}, are capable of capturing non-equilibrium effects and have been used to study fluid behaviors at the nanoscale. Particularly, the DUGKS method exhibits the nice unified-preserving property to enable it a reliable method for simulating multiscale flows \cite{PhysRevE.107.025301}, and has been adopted to study
some typical nano-confined fluid flows \cite{shan2021contribution,shan2020discrete,shan2021pore}. However, the kinetic model \cite{guo2005simple}, based on
which the nano-confined DUGKS \cite{shan2020discrete} was developed, involves a number of spatial integrals associated with average density and interaction forces, which lead to severe computational challenges. Actually, previous applications of this DUGKS are limited to one-dimensional problems.

In this study, we develop an improved DUGKS by evaluating the spatial integrals with much reduced grid points without sacrificing the accuracy, thus the computational efficiency is significantly improved. The remainder of this paper is organized as follows. Section ~\ref{sec:2} provides a review of the strongly inhomogeneous kinetic model at the nanoscale. Next, the original DUGKS for the kinetic model is reviewed in Section ~\ref{sec:3}. In Section ~\ref{sec:4}, we present the detailed formulations of the proposed DUGKS, including the treatments of the integrals. In Section ~\ref{sec:5}, we assess the performance of the proposed DUGKS through several numerical tests. Section ~\ref{sec:6} investigates two-dimensional flows. Finally, concluding remarks are presented in Section ~\ref{sec:7}.

\section{Kinetic model for strongly inhomogeneous fluids}\label{sec:2}
The DUGKS method \cite{shan2020discrete} is based on the tractable kinetic model for nano-confined strongly inhomogeneous flows \cite{guo2005simple}
\begin{equation}
\frac{\partial f}{\partial t} + \bm{\xi} \cdot \nabla f + m^{-1} \left[ \bm{F} - \nabla \left( \phi_{w} + \phi_{m} \right) \right] \cdot \nabla_{\bm{\xi}} f = -\frac{f-f^{eq}}{\tau}+J_{ex},
\label{eq:1}
\end{equation}
\noindent where $f=f(\bm{r},\bm{\xi},t)$ represents the distribution function of molecules moving with velocity $\bm{\xi}$ at position $\bm{r}$ and time $t$, $m$ denotes the molecular mass, $\bm{F}$ is the external force acting on the molecules, $\phi_w$ refers to the wall potential, and $\phi_m$ indicates the potential arising from long-range interactions. The right hand side of Eq. \eqref{eq:1} represents the overall collision effects. The Bhatnagar-Gross-Krook (BGK) relaxation term \cite{PhysRev.94.511} describes the local collision, where $\tau$ is the relaxation time and $f^{eq}$ is the Maxwellian equilibrium distribution function
\begin{equation}
f^{eq}=n\left(\frac{1}{2 \pi R T}\right)^{3 / 2} \exp \left[-\frac{ (\boldsymbol{\xi}-\boldsymbol{u})^2}{2 R T}\right],
\label{eq:2}
\end{equation}
with $R$ being the specific gas constant and $T$ the temperature. The number density $n$ and velocity $\bm{u}$ can be obtained by taking the moments of the distribution function as
\begin{equation}
    n = \int f \, d\bm{\xi}, \quad \bm{u}=n^{-1}\int \bm{\xi}f \, d\bm{\xi}.
    \label{eq:3}
\end{equation}

The term $J_{ex}$ on the right hand side of Eq. \eqref{eq:1} describes the nonlocal collision of dense gas 
\begin{equation}
J_{ex}=-V_0f^{eq}(\bm\xi-\bm{u})\cdot[2\bm{A}(\bar{n})\chi(\bar{n})+\bm{B}(\bar{n})\bar{n}],
\label{eq:4}
\end{equation}
where $\bar{n}(\bm{r})=\int \omega(\bm{r}'){n}(\bm{r} + \bm{r}') \, d\bm{r}'$ is the average density, $V_0=2\pi\sigma^3/3$ is the second-order virial coefficient, and $\chi$ is the pair correlation function(PCF), which can be evaluated by the average density according to the Carnahan and Starling equation of state \cite{carnahan1969equation}
\begin{equation}
\chi=\frac{1-0.5\eta}{(1-\eta)^3}, \quad \eta=\frac{\bar{n}\pi\sigma^3}{6} \text {. }
\label{eq:5}
\end{equation}

The two vector operators $\bm{A}$ and $\bm{B}$ are the generalizations of the gradients of density and PCF \cite{guo2005simple}
\begin{equation}
\begin{aligned}
& \bm{A}=\frac{120}{\pi\sigma^5} \int_{|\bm{r}'|<\sigma/2} \bm{r}' \bar{n}(\bm{r} + \bm{r}') \, d\bm{r}', \\
& \bm{B}=\frac{120}{\pi\sigma^5} \int_{|\bm{r}'|<\sigma/2} \bm{r}' \chi(\bm{r} + \bm{r}') \, d\bm{r}'.
\end{aligned}
\label{eq:6}
\end{equation}
They measure the first spatial moment (i.e., the imbalance) of $\bar{n}$ and $\chi$ around $\boldsymbol{r}$ over the neighbor $|\boldsymbol{r}-\boldsymbol{r}'|<\sigma/2$, and can be regarded as nonlocal (or weighted) gradients. It can be shown that for a weakly inhomogeneous fluid, i.e., $\bar{n}=n$, $A$ and $\bm{B}$ reduce to $\nabla n$ and $\nabla \chi$. 

The relation time $\tau$ in the BGK collision term is determined by \cite{heidaryan2010novel}
\begin{equation}
\tau = \frac{\mu}{p},\quad p=Znk_BT
\label{eq:7}
\end{equation}
where $p$ is the pressure, $Z$ denotes the compressibility factor, $k_B$ is the Boltzmann constant, and $\mu$ is the dynamic viscosity of dense gas \cite{chapman1990mathematical}
\begin{equation}
    \mu=1.016\frac{5}{16\sigma^2}\sqrt\frac{mk_BT}{\pi}\bar{n}V_0\left(\frac{1}{\bar{n}V_0\chi}+0.8+0.7614\bar{n}V_0\chi\right).
    \label{eq:8}
\end{equation}

In the forcing term of Eq. \eqref{eq:1}, the long-range attractive potential $\phi_m$ is determined by the mean-field theory \cite{grmela1971kinetic}
\begin{equation}
\phi_m(\bm{r})=\int n({\bm{r}+\bm{r}'})\phi_{att}(|\bm{r}'|) \, d\bm{r}',
\label{eq:9}
\end{equation}
\noindent where $\phi_{att}$ represents the attractive part of the interaction potential between fluid molecules. For the 12-6 Lennard-Jones (LJ) potential, $\phi_{att}$ is given by \cite{rowley1975monte}
\begin{equation}
\phi_{\mathrm{att}}(|\boldsymbol{r}'|) = 
\begin{cases}
0, & |\boldsymbol{r}'| \leq \sigma \\
-4\epsilon \left( \frac{\sigma}{|\boldsymbol{r}'|} \right)^6, & |\bm{r}'|>\sigma
\end{cases},
\label{eq:10}
\end{equation}
\noindent where $\sigma$ and $\epsilon$ are the range and energy parameters of the potential.

The wall potential $\phi_w$ represents the fluid-solid molecular interactions. For example, if the interaction between a pair of fluid and wall molecules is modeled by the 12-6 LJ potential, the potential experienced by a fluid molecule arising from the cumulative effect of all molecules in a flat solid wall can be described by the 10-4-3 LJ potential \cite{steele1973physical}
\begin{equation}
\phi(y)=2 \pi n_s \epsilon_{wf} \sigma_{wf}^2\left[\frac{2}{5}\left(\frac{\sigma_{w f}}{y}\right)^{10}-\left(\frac{\sigma_{w f}}{y}\right)^4-\frac{\sigma_{w f}{ }^4}{3 \Delta(y+0.61 \Delta)^3}\right], \quad \Delta=\sigma_{w f} / \sqrt{2} \text {, }
\label{eq:11}
\end{equation}
\noindent where $y$ is the distance to the wall, $n_s$ is the number density of solid molecules, and $\epsilon_{wf}$ and $\sigma_{wf}$ are the energy and length parameters of the fluid-solid molecular interactions, respectively.

\section{Original DUGKS for strongly inhomogeneous kinetic model}\label{sec:3}
\subsection{Original DUGKS}
The original DUGKS \cite{shan2020discrete} solves Eq. \eqref{eq:1} via the Strang splitting strategy. In this section, we will give a brief description of the method. First, the kinetic equation \eqref{eq:1} is rewritten as  
\begin{equation}
\frac{\partial f}{\partial t} + \bm{\xi} \cdot \nabla f = Q + {G}.
\label{eq:12}
\end{equation}
where $Q$ is the BGK relaxation term and ${G}$ represents contribution from total force
\begin{equation}
{G}=-m^{-1} \left[ \bm{F} - \nabla \left( \phi_{w} + \phi_{m} \right) \right] \cdot \nabla_{\bm{\xi}} f + J_{ex}\approx\boldsymbol{a}\cdot\frac{\boldsymbol{\xi}-\boldsymbol{u}}{RT}f^{eq},
\label{eq:13}
\end{equation}
with $\boldsymbol{a}=\boldsymbol{F}-\nabla (\phi_w+ \phi_m)-V_0RT(2\boldsymbol{A}\chi+\boldsymbol{B}\bar{n})$ being the total force. Then the DUGKS solves Eq. \eqref{eq:12} by discretizing the following three time-splitting equations successively at each time step
\begin{subequations}\label{eq:14}
    \begin{equation}
        \partial_t f = 0.5G,
        \label{eq:14a}
    \end{equation}
    \begin{equation}
        \frac{\partial f}{\partial t} + \xi \cdot \nabla f = Q,
        \label{eq:14b}
    \end{equation}
    \begin{equation}
        \partial_t f = 0.5G,
        \label{eq:14c}
    \end{equation}
\end{subequations}
which are called the pre-forcing step, standard DUGKS step, and post-forcing step, respectively.

We now present the DUGKS step for Eq. \eqref{eq:14b}. The physical space is first discretized into a set of cells with finite-volume scheme. Then we integrate Eq. \eqref{eq:14b} over the cell $V_m$ centered at $\bm{r}_m$ from time $t_n$ to $t_{n+1}$, which leads to
\begin{equation}
f_m^{n+1}-f_m^n+\frac{\Delta t}{\left|V_m\right|} F^{n+1 / 2}=\frac{\Delta t}{2}\left(Q_{m}^{n+1}-Q_{m}^n\right),
\label{eq:15}
\end{equation}
where $\Delta t= t_{n+1}-t_n$ is the time step, $(.)^n_m$ denotes the volume-averaged value of the cell, and
\begin{equation}
F^{n+1/2}=\int_{\partial V_{m}} (\bm\xi \cdot \bm{n}) f ( \bm{r}, \bm{\xi}, t_{n}+h) \, d\bm{S},
\label{eq:16}
\end{equation}
is the microscopic flux across the cell interface, with $|V_m|$ and $\partial V_m$ being the volume and surface of cell $V_m$, $\bm{n}$ the outward unit vector normal to cell interface, and $h=\Delta t/2$ the half time step. 

The implicitness of Eq. \eqref{eq:15} can be removed by introducing two new distribution functions $\tilde f$ and $\tilde f^+$ as
\begin{equation}
\begin{aligned}
& \tilde f=f-\frac{\Delta t}{2}Q=\frac{2\tau+\Delta t}{2\tau}f-\frac{\Delta t}{2\tau}f^{eq}, \\
& \tilde f^+=f+\frac{\Delta t}{2}Q=\frac{2\tau-\Delta t}{2\tau+\Delta t}f+\frac{2\Delta t}{2\tau+\Delta t}f^{eq}. 
\end{aligned}
\label{eq:17}
\end{equation}
Then Eq. \eqref{eq:15} can be rewritten as
\begin{equation}
\tilde f^{n+1}_{m}=\tilde f^{+,n}_{m}-\frac{\Delta t}{V_{m}}F^{n+1/2}.
\label{eq:18}
\end{equation}
Since the BGK collision operator conserves mass and momentum, the number density and velocity can be calculated as 
\begin{equation}
    n = \int \tilde f \, d\bm{\xi}, \quad \bm{u}=n^{-1}\int \bm{\xi} \tilde f \, d\bm{\xi}.
    \label{eq:19}
\end{equation}
Therefore, we can track the new distribution function $\tilde f$ instead of the original one in the DUGKS.

The key point in updating $\tilde f$ according to Eq. \eqref{eq:18} is to calculate the flux $F^{n+1/2}$, which requires the reconstruction of the original distribution function $f$ at time $t_{n}+h$ on the cell interface. To this end, we integrate Eq. \eqref{eq:14b} with a half time step along the characteristic line with end point at the cell interface $\boldsymbol{r}_b$
\begin{equation}
f\left(\boldsymbol{r}_b, \boldsymbol{\xi}, t_n+h\right)-f\left(\boldsymbol{r}_b-\boldsymbol{\xi} h, \boldsymbol{\xi}, t_n\right)=\frac{h}{2}\left[{Q}\left(\boldsymbol{r}_b, \boldsymbol{\xi}, t_n+h\right)+{Q}\left(\boldsymbol{r}_b-\boldsymbol{\xi} h, \boldsymbol{\xi}, t_n\right)\right].
\label{eq:20}
\end{equation}
To remove the implicitness, again we introduce two auxiliary distribution functions $\bar{f}=f-\frac{h}{2}Q$ and $\bar{f}=f+\frac{h}{2}Q$, and then Eq. \eqref{eq:20} can be rewritten as
\begin{equation}
\bar{f}\left(\boldsymbol{r}_b, \boldsymbol{\xi}, t_n+h\right)=\bar{f}^{+}\left(\boldsymbol{r}_b-\boldsymbol{\xi} h, \boldsymbol{\xi}, t_n\right).
\label{eq:21}
\end{equation}
By assuming $\bar{f}^{+}\left(\boldsymbol{r}_b-\boldsymbol{\xi} h, \boldsymbol{\xi}, t_n\right)$ is linear in the cell $V_m$, we can reconstruct it as
\begin{equation}
\bar{f}^{+}\left(\boldsymbol{r}_b-\boldsymbol{\xi} h, \boldsymbol{\xi}, t_n\right)=\bar{f}^{+}\left(\boldsymbol{r}_m, \boldsymbol{\xi}, t_n\right)+(\boldsymbol{r}_b-\boldsymbol{\xi} h-\boldsymbol{r}_m) \cdot \boldsymbol{\sigma}_m, \quad (\boldsymbol{r}_b-\boldsymbol{\xi}h) \in V_m,
\label{eq:22}
\end{equation}
where $\boldsymbol{\sigma}_m$ is the slope of $\bar{f}^+$ in the cell. For nano-confined fluid systems, the slope $\boldsymbol{\sigma}_m$ can be calculated using the van Leer limiter \cite{van1977towards}. Taking the component in the $y$ direction as an example, the slope can be written as
\begin{equation}
\sigma_{m, y}=\left[\operatorname{sign}\left(s_1\right)+\operatorname{sign}\left(s_2\right)\right] \frac{\left|s_1\right|\left|s_2\right|}{\left|s_1\right|+\left|s_2\right|},
\label{eq:23}
\end{equation}
where
\begin{equation}
s_1=\frac{\bar{f}^{+}\left(x_{i},y_j,z_k\right)-\bar{f}^{+}\left(x_{i},y_{j-1},z_k\right)}{y_j-y_{j-1}}, \quad s_2=\frac{\bar{f}^{+}\left(x_{i},y_{j+1},z_k\right)-\bar{f}^{+}\left(x_i,y_j,z_k\right)}{y_{j+1}-y_j} .
\label{eq:24}
\end{equation}

After obtaining $\bar{f}\left(\boldsymbol{r}_b, \boldsymbol{\xi}, t_n\right)$, the conserved variables can be computed by taking the moments of $\bar{f}\left(\boldsymbol{r}_b, \boldsymbol{\xi}, t_n+h\right)$. As a result, the equilibrium distribution function can be determined and the original distribution can be obtained
\begin{equation}
f\left(\boldsymbol{r}_b, \boldsymbol{\xi}, t_n+h\right)=\frac{2 \tau}{2 \tau+h} \bar{f}\left(\boldsymbol{r}_b, \boldsymbol{\xi}, t_n+h\right)+\frac{h}{2 \tau+h} f^{\mathrm{eq}}\left(\boldsymbol{r}_b, \boldsymbol{\xi}, t_n+h\right) .
\label{eq:25}
\end{equation}
Then the flux $F^{n+1/2}$ can be obtained according to Eq. \eqref{eq:16} and the DUGKS step for Eq. \eqref{eq:14b} can be completed by tracking the new distribution function $\tilde f$. 

In the presence of the force term $G$, the pre-forcing step is implemented by integrating Eq. \eqref{eq:14a} using a backward Euler scheme over a time step $\Delta t$
\begin{equation}
f^*=f+\frac{\Delta t}{2}{G}(n^*,\boldsymbol{u}^*),
\label{eq:26}
\end{equation}
where the number density $n^*$ and velocity $\boldsymbol{u}^*$ in the forcing term are obtained by taking the moments of Eq. \eqref{eq:26} as
\begin{equation}
n^*=n, \quad \boldsymbol{u}^*=\boldsymbol{u}+\frac{\Delta t}{2}{\boldsymbol{a}}.
\label{eq:27}
\end{equation}
According to Eqs. \eqref{eq:17} and \eqref{eq:26}, we can get
\begin{equation}
\begin{aligned}
\tilde{f}^*=\tilde{f}+\frac{\Delta t}{2 \tau}\left[f^{e q}(n, \boldsymbol{u})-f^{e q}\left(n^*, \boldsymbol{u}^*\right)\right]+\frac{(2 \tau+\Delta t) \Delta t}{4 \tau} {G}(n, \boldsymbol{u}) .
\label{eq:28}
\end{aligned}
\end{equation}
The post-forcing step can be implemented similarly. It is noted that with the Strang splitting approach, the scheme is of second-order accuracy in time.

In the above DUGKS, a number of spatial integrals are required to be evaluated numerically, including the mean-field attractive potential $\phi_m$, average density $\bar{n}$, and the nonlocal gradient operators $\boldsymbol{A}$ and $\boldsymbol{B}$. In the original DUGKS algorithm, the integrals are evaluated by counting all mesh points in the integration domain, which is of order $N_\sigma$. These computations lead to a significant computational cost $O(NN_\sigma)$, where $N$ is the total number of mesh cells and $N_\sigma = N_{\sigma x}N_{\sigma y}N_{\sigma z}$ is the number of mesh points in a cubic domain with edge length $\sigma$. Actually, previous applications were limited to one-dimensional cases \cite{shan2021contribution,shan2020discrete,shan2021pore}.

\subsection{Boundary Condition} 
Both fluid-fluid and fluid-solid molecular interactions become significant at the nanoscale, resulting in surface diffusion and gas adsorption at the boundary. Therefore, bounce-back and diffuse boundary conditions become ineffective. To accurately describe fluid-fluid and fluid-solid molecular interactions, a bounce-back boundary condition with slip velocity is employed \cite{shan2022molecular} 
\begin{equation}
    f(\bm{r}_{w}, \bm{\xi}_{i}, t) = f(\bm{r}_{w}, -\bm{\xi}_{i}, t) + 2\rho_{w}W_{i} \frac{\bm{\xi}_{i} \cdot \bm{u}_{s}}{RT}, \quad \bm{\xi}_{i} \cdot \bm{n} > 0
    \label{eq:29}
\end{equation}
where $\bm{n}$ denotes the unit vector normal to the wall pointing to the fluid, $\rho_w$ represents the fluid density at the wall, $W_i$ is the weight coefficient associated with the discrete velocity $\bm\xi_i$, and $\bm{u}_s$ is the slip velocity induced by intermolecular forces, which can be expressed as
\begin{equation}
u_s=f_{w e t} \delta \frac{2 k_B T}{h} \exp \left(-\frac{a \epsilon_{w f}}{k_B T}\right) \sinh \left(\frac{\gamma S \delta}{2 k_B T}\right),
\label{eq:30}
\end{equation}
where $f_{wet}$ is a wettability factor, $\delta$ is the hopping length, $h$ is the hopping rate, $a$ is a constant characterizing the strength of fluid-solid interactions, $\gamma$ is the shear stress, and $S$ is the effective area that the fluid molecule experiences shear.

\subsection{Algorithm} 
Given the distribution function $\tilde f$, the number density $n$ and the velocity $\boldsymbol{u}$ at the $n$-th iteration step in each cell, their values at the $(n+1)$-th step are updated as follows:

\noindent Step 1. Calculate the average density $\bar{n}$ and PCF $\chi$ in Eq. \eqref{eq:4} and then evaluate the gradients of the average density $\bar{n}$ and PCF $\chi$ based on Eq. \eqref{eq:6}.

\noindent Step 2. Evaluate the long-range attractive potential $\phi_{m}$ in Eq. \eqref{eq:9}.

\noindent Step 3. Couple the force term ${G}$ according to the Strang splitting method \eqref{eq:28}.

\noindent Step 4. Calculate the distribution function $\bar{f}^+$ and its gradient in Eq. \eqref{eq:22}, and evaluate $\bar{f}(\boldsymbol{r}_b)$ using Eqs. \eqref{eq:21}, then compute the original distribution function $f(\boldsymbol{r}_b)$ based on Eq. \eqref{eq:25}. 

\noindent Step 5. Calculate the flux $\boldsymbol{F}^{n+1/2}$ in Eq. \eqref{eq:16}.

\noindent Step 6. Calculate the distribution function $\tilde f^+$ and update the distribution function $\tilde f$ based on Eq. \eqref{eq:18}.

\noindent Step 7. Repeat Step 1 to Step 3 to complete the post-forcing step.

\noindent Step 8. Repeat the above steps until the following convergence criteria are satisfied:
\begin{equation}
    \sqrt{\frac{\sum\limits_{m}(n_m^{l+1000}-n_m^l)^2}{\sum\limits_{m}{(n_m^l)^2}}}<10^{-6} \quad \text{and} \quad \sqrt{\frac{\sum\limits_{m}(\bm{u}_m^{l+1000}-\bm{u}_m^l)^2}{\sum\limits_{m}{(\bm{u}_m^l)^2}}}<10^{-6},
    \label{eq:31}
\end{equation}
where $(.)_m^l$ denotes the approximate value of a cell $m$ at the $l$-th iteration step.

\section{An improved DUGKS with simplified quadratures}\label{sec:4}
To reduce computational cost, some more efficient strategies are developed to evaluate the integrals without sacrificing numerical accuracy.

\subsection{Mean-field attractive potential}
In both the original DUGKS and the improved DUGKS, the physical domain is discretized into a set of mesh cells. For simplicity without loss of generality, we assume a uniform Cartesian mesh in what
follows. In the original DUGKS, the mean-field attractive potential at $\boldsymbol{r}$ is computed by sampling all mesh points in the integration domain
\begin{equation}
\phi_m(\bm{r})=\sum\limits_{m} W_m n(\bm{r} + \bm{r}_m') \phi_{att}(|\bm{r}_m'|) \, , \quad |\bm{r}_m'|<3\sigma
\label{eq:32}
\end{equation}
where the weight $W_m=|V_m|$ is the volume of cell $V_m$ centered at $\boldsymbol{r}+\boldsymbol{r}'_m$. This direct computation results in $O(NN_\sigma)$ operations. For nano-confined fluid systems, fine mesh resolution is usually required to capture the strong inhomogeneity, and thus leads to intensive computational demands.

To compute the mean-field potential efficiently, we introduce a numerical scheme with a coarse sampling set. Note that $\phi_{att}(|\bm{r}'|)$ in Eq. \eqref{eq:10} is smooth enough and decays to zero rapidly with increasing $|\bm{r}'|$. Therefore, as the distance $|\bm{r}'|$ increases, fewer quadrature points are required to evaluate the attractive potential
\begin{equation}
\phi_m(\bm{r})=\sum\limits_{m} W_m n(\bm{r} + \bm{r}_m')\phi_{att}(|\bm{r}_m'|) \, , \quad |\bm{r}_m'|<3\sigma
\label{eq:33}
\end{equation}
where $\bm{r}'_m=a(\frac{i^2-1}{2},\frac{j^2-1}{2},\frac{k^2-1}{2})$ (with $i,j,k \in \mathbb{Z}^+ = \{1, 2, 3, \dots\} $) denotes the set of sampling points used to evaluate the mean-field potential, $W_m=ijka^3$ is the associated weight and $a$ measures the sampling spacing. With the midpoint rule, the above scheme achieves second-order spatial accuracy, consistent with DUGKS. The integral can be accurately calculated according to Eq.~\eqref{eq:33} with far fewer sampling points than the direct computation using all mesh points in the integration domain, resulting in linear complexity $O(N)$. Generally, $a$ can take a small value for strongly inhomogeneous systems and a larger value for weakly inhomogeneous systems.

\subsection{Average density}
In the kinetic model \cite{guo2005simple}, the average density $\bar{n}(\boldsymbol{r})$ is defined as a spatially weighted average of the local density with a specific weighting function $\omega(\boldsymbol{r}')$ \cite{vanderlick1989molecular}
\begin{equation}
\bar{n}(\bm{r})=\int \omega(\bm{r}'){n}(\bm{r} + \bm{r}') \, d\bm{r}'.
\label{eq:34}
\end{equation}
A simple and intuitive approach is the volumn-averaging method (VAM), which averages the local density within a sphere of radius $\sigma$ \cite{nordholm1980generalized}
\begin{equation}
\omega(\bm{r}')= H(\sigma-|\bm{r}'|) \frac{3}{4 \pi \sigma^3},
\label{eq:35}
\end{equation}
where $H$ denotes the Heaviside step function. Although the VAM can prevent the average density from exceeding the close-packing limit, it fails to satisfy the normalization condition near the wall and may become ineffective at high mean densities. To address this, Tarazona et al. \cite{tarazona1985free} proposed a weighting function depending on both relative position and average density
\begin{equation}
\omega(\bm{r}')=\omega_0(\bm{r}')+\omega_1(\bm{r}')\bar{n}(\bm{r} + \bm{r}')+\omega_2(\bm{r}')\bar{n}(\bm{r} + \bm{r}')^2,
\label{eq:36}
\end{equation}
which is adopted in the original DUGKS. The density-independent weighting functions are given as
\begin{equation}
\begin{aligned}
& \omega_1(|\boldsymbol{r}'|)= \begin{cases}\frac{3}{4 \pi \sigma^3}, & |\boldsymbol{r}'| \leqslant \sigma, \\
0, & |\boldsymbol{r}'|>\sigma,\end{cases} \\
& \omega_2(|\boldsymbol{r}'|)= \begin{cases}0.475-0.648\left(\frac{|\boldsymbol{r}'|}{\sigma}\right)+0.113\left(\frac{|\boldsymbol{r}'|}{\sigma}\right)^2, & |\boldsymbol{r}'| \leqslant \sigma, \\
0.288\left(\frac{\sigma}{|\boldsymbol{r}'|}\right)-0.924+0.764\left(\frac{|\boldsymbol{r}'|}{\sigma}\right)-0.187\left(\frac{|\boldsymbol{r}'|}{\sigma}\right)^2, & \sigma<|\boldsymbol{r}'| \leqslant 2 \sigma, \\
0, & |\boldsymbol{r}|>2 \sigma,\end{cases} \\
& \omega_3(|\boldsymbol{r}'|)= \begin{cases}\frac{5 \pi \sigma^3}{144}\left[6-12\left(\frac{|\boldsymbol{r}'|}{\sigma}\right)+5\left(\frac{|\boldsymbol{r}'|}{\sigma}\right)^2\right], & |\boldsymbol{r}'| \leqslant \sigma, \\
0, & |\boldsymbol{r}'|>\sigma .\end{cases}
\end{aligned}
\label{eq:37}
\end{equation}
The spatial averaging range in Tarazona's method is defined as a spherical domain with a diameter of $4\sigma$. As a result, this method may become inaccurate when the dimension is less than $4\sigma$. Furthermore, the presence of multiple weighting functions increases the computational cost. 

The strength and range of intermolecular interactions vary across different fluid-solid systems, resulting in different density inhomogeneity. Accordingly, the averaging range for the density should also differ. The density variations may be obscured if the averaging range is too large, whereas a small range may result in density oscillations. Therefore, we propose an adaptive volume-averaging method (AVAM) in which the averaging range varies according to the characteristics of the fluid-solid system and consistently satisfies the normalization condition. Specifically, the weighting function in AVAM is defined by
\begin{equation}
\omega(\bm{r}')=\frac{H(d-|\bm{r}'|) }{\int H(d-|\bm{r}'| \, d\bm{r}'},
\label{eq:38}
\end{equation}
with
\begin{equation}
d=k(\frac{T_r\epsilon}{n_0\epsilon_{wf}})^{1/3},
\label{eq:39}
\end{equation}
where $d$ is the averaging range for the local density, $T_r=k_BT/\epsilon_{ff}$ is the reduced temperature, $n_0$ is the average number density, and $k$ is a coefficient related to the fluid-solid system. It can be seen that as ${(T_r\epsilon)}/{(n_0\epsilon_{wf})}$ decreases, the density inhomogeneity becomes more pronounced, leading to a smaller averaging range for density. 

To reduce computational cost, the average density is computed on a coarse scale
\begin{equation}
\bar{n}(\bm{r})={h_\sigma^3}\sum\limits_{m} \omega(\bm{r}_m') {n}(\bm{r} + \bm{r}_m') ,
\label{eq:40}
\end{equation}
with $\bm{r}_m'=(i-1,j-1,k-1)h_\sigma$ being the sampling offsets and $h_\sigma=\sigma/10$ the sampling spacing in each direction. It can be shown that the scheme is second-order accurate in space, in line with the accuracy of DUGKS. To capture strong inhomogeneity at the nanoscale, the cell spacing in DUGKS is much smaller than $\sigma/10$. Therefore, the computational cost can be significantly reduced by employing Eq. \eqref{eq:40}.

\subsection{Nonlocal gradient}
In order to efficiently calculate nonlocal gradients $\boldsymbol{A}$ and $\boldsymbol{B}$, we propose a numerical scheme based on finite-difference method. In what follows, we only present the procedure for computing $\boldsymbol{A}$ and the method can be applied to the calculation of $\boldsymbol{B}$. Expanding $\bar{n}(\boldsymbol{r}+\boldsymbol{r}')$ in the integral of Eq. \eqref{eq:6} into a Taylor series around the mesh point $\bm{r}=(x_i,y_j,z_k)$ and retaining terms up to fourth order, we have
\begin{equation}
\bm{A}=\nabla\bar{n}+\frac{20}{\pi\sigma^5} \int_{|\bm{r}'|<\sigma/2} \bm{r}' (\bm{r}'\cdot\nabla)^3 \bar{n} \, d\bm{r}'.
\label{eq:41}
\end{equation}
For convenience, we take the $y$-direction as an example in subsequent derivations. The component of $A$ in the $y$-direction is
\begin{equation}
    \bm{A}_y=\frac{\partial\bar{n}}{\partial{y}}+\frac{\sigma^2}{56}(\frac{\partial^3\bar{n}}{\partial{x}\partial{x}\partial{y}}+\frac{\partial^3\bar{n}}{\partial{y}\partial{y}\partial{y}}+\frac{\partial^3\bar{n}}{\partial{y}\partial{z}\partial{z}}).
\label{eq:42}
\end{equation}

We define two finite-difference operators $\bar{\nabla}_1(.)$ and $\bar{\nabla}_2(.)$, in which the expression along the $y$-direction is
\begin{equation}
\begin{aligned}
\bar{\nabla}_{1y}\bar{n}=& \frac{\bar{n}(x_i,y_j+a,z_k)-\bar{n}(x_i,y_j-a,z_k)}{2h}, \\
\bar{\nabla}_{2y}\bar{n}=& \frac{1}{4}{\left[\frac{\bar{n}\left(x_i+a, y_j+a, z_k+a\right)-\bar{n}\left(x_i+a, y_j-a, z_k+a\right)}{2 a}\right.} \\ & +\frac{\bar{n}\left(x_i+a, y_j+a, z_k-a\right)-\bar{n}\left(x_i+a, y_j-a, z_k-a\right)}{2 a} \\ & +\frac{\bar{n}\left(x_i-a, y_j+a, z_k+a\right)-\bar{n}\left(x_i-a, y_j-a, z_k+a\right)}{2 a} \\ & \left.+\frac{\bar{n}\left(x_i-a, y_j+a, z_k-a\right)-\bar{n}\left(x_i-a, y_j-a, z_k-a\right)}{2 a}\right],
\end{aligned}
\label{eq:43}
\end{equation}
where $a$ represents the distance from reference points to $\bm{r}$ in each direction. Then we get a hybrid finite-difference scheme $\bar{\nabla}$ by combining $\bar{\nabla}_1$ and $\bar{\nabla}_2$
\begin{equation}
\bar{\nabla}\bar{n}=\frac{2}{3}\bar\nabla_1\bar{n}+\frac{1}{3}\bar\nabla_2\bar{n}. 
\label{eq:44}
\end{equation}
If we take $a=\sqrt{\frac{6}{56}}\sigma$, then the Taylor expansion of $\bar\nabla\bar{n}$ around $\bm{r}$ give that
\begin{equation}
    \bar{\nabla}_y\bar{n}=\frac{\partial\bar{n}}{\partial{y}}+\frac{\sigma^2}{56}(\frac{\partial^3\bar{n}}{\partial{x}\partial{x}\partial{y}}+\frac{\partial^3\bar{n}}{\partial{y}\partial{y}\partial{y}}+\frac{\partial^3\bar{n}}{\partial{y}\partial{z}\partial{z}})+O(a^5),
\label{eq:45}
\end{equation}
which is exactly an approximation of Eq. \eqref{eq:42}. Particularly, using Eq. \eqref{eq:45} to evaluate the nonlocal gradient $\boldsymbol{A}$ reduces the computational cost to $O(N)$. Similarly, we can compute the vector operator $\boldsymbol{B}$.

\begin{figure}[ht]
     \centering
     \includegraphics[width=0.4\textwidth]{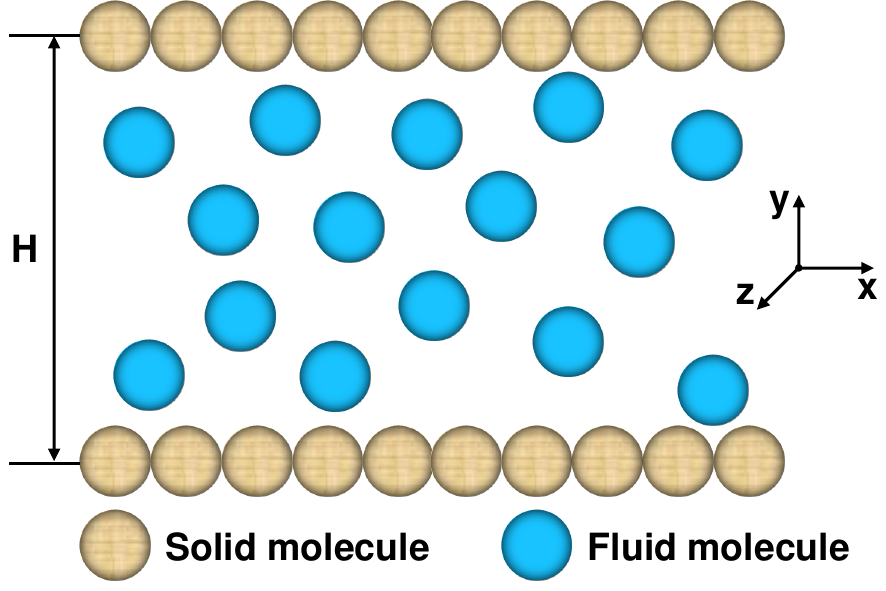}
     \caption{Schematic of fluid molecules confined between two parallel solid walls.}
     \label{FIG1}
\end{figure}

\section{Numerical performance tests}\label{sec:5}
In this section, the performance of the proposed DUGKS is assessed through two fundamental problems: the static structure of inhomogeneous fluids and the dynamics of force-driven flows between two parallel plates, as shown in~\cref{FIG1}. The weighting functions used to compute the average density can be reduced to one- and two-dimensional forms via integration (see Appendix A for details). A half-range Gauss-Hermit discrete velocity set with $8\times8$ velocities is used to capture the non-equilibrium effects at the nanoscale \cite{shizgal1981gaussian}.

\begin{figure}[H]
  \centering

  \subfloat{\label{f2:sub1}
    \includegraphics[width=8.5cm]{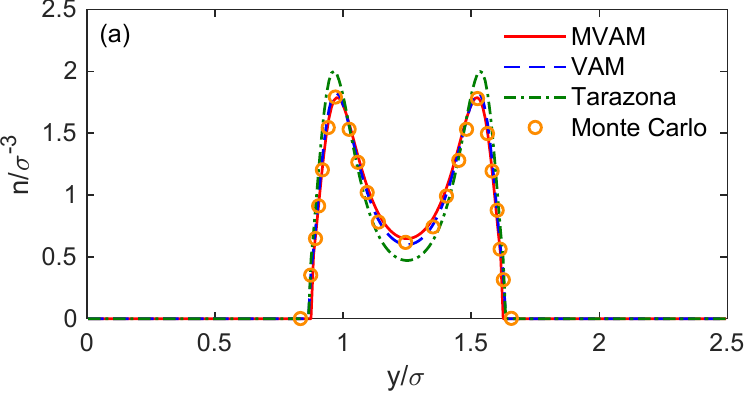}}
    
    \subfloat{\label{f2:sub2}
    \includegraphics[width=8cm]{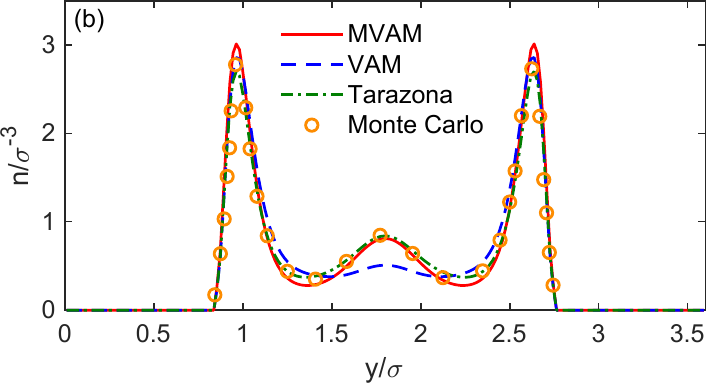}}

      \subfloat{\label{f2:sub3}
    \includegraphics[width=8cm]{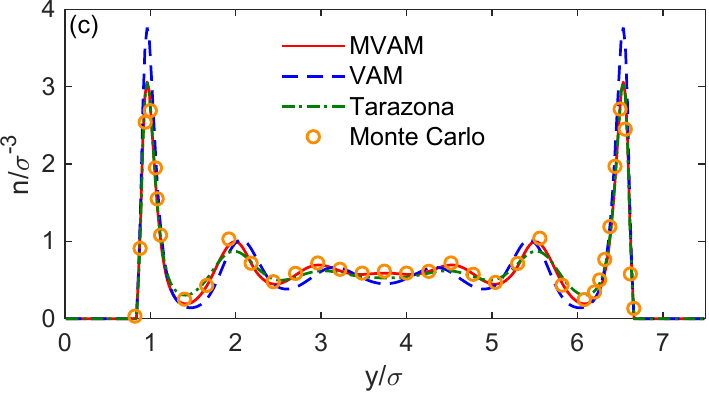}}

  \caption{Density distributions of LJ fluids at different widths and average densities. (a) $H=2.5\sigma$, $n_0=0.335\sigma^{-3}$; (b) $H=3.6\sigma$, $n_0=0.476\sigma^{-3}$; (c) $H=7.5\sigma$, $n_0=0.561\sigma^{-3}$. Monte Carlo results are taken from Ref. \cite{snook1980solvation}.}
  \label{FIG2}
\end{figure}

\subsection{Static structure}
\subsubsection{Verification of the accuracy of AVAM }
The density distributions of LJ fluids between two parallel plates with channel widths \(H = 2.5\sigma\), \(3.6\sigma\), and \(7.5\sigma\) are first investigated. The corresponding average number densities are \({n_0} = 0.335\sigma^{-3}\), \(0.476\sigma^{-3}\), and \(0.551\sigma^{-3}\), respectively. For all cases, we set $T_r=1.2$ and $\epsilon=\epsilon_{wf}$. The coefficient $k$ in Eq. \eqref{eq:39} is set to be 1.5. To capture density inhomogeneity, the physical domain is discretized by a mesh with $N_{\sigma y}=100$ \cite{shan2021contribution}, where $N_{\sigma y}$ represents the number of cells within a molecular diameter $\sigma$ along the $y$-direction. To facilitate the verification of the accuracy of AVAM, the integrals in the kinetic model \eqref{eq:1} are computed by sampling all mesh points within the integration domain, as done in the original DUGKS. For comparison, the results in Ref.~\cite{snook1980solvation} obtained using the Monte Carlo method are adopted as reference solutions. 

The density distributions across the channel obtained using different weighting functions are shown in~\cref{FIG2}. It can be observed that the results of Tarazona's method agree well with reference solutions as $H=3.6\sigma$ and $H=7.5\sigma$. However, the discrepancies are not negligible as $H=2.5\sigma$ because the channel width is significantly smaller than the averaging range of Tarazona's method. While the density distribution obtained from the VAM compares well with the reference solution at $n_0=0.335\sigma^{-3}$, the density variation can be observed as $n_0$ increases. This is due to the fact that density inhomogeneity increases with $n_0$, which leads to a varying averaging range. Good agreement between the AVAM and Monte Carlo method is observed for three cases. These numerical comparisons clearly demonstrate the accuracy of the proposed AVAM. 

\begin{figure}[H]
  \centering
  
    \subfloat{\label{f3:sub1}
    \includegraphics[width=0.46\textwidth]{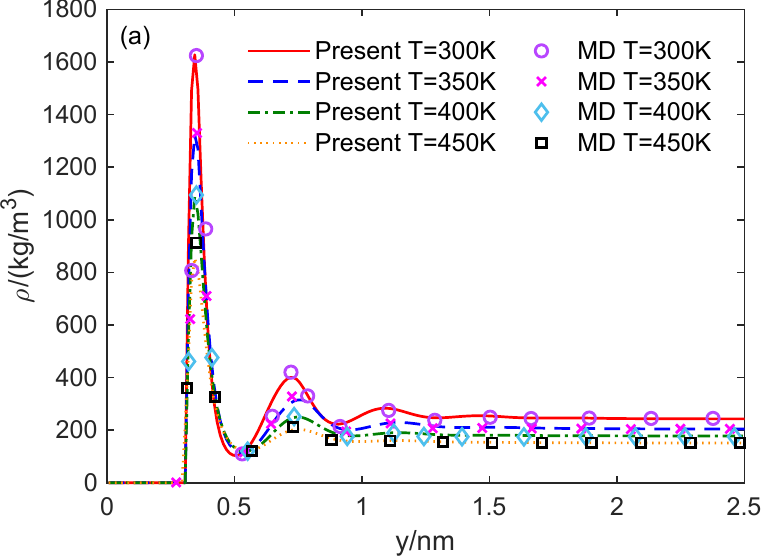}}
    \hspace{0.02\textwidth}
    \subfloat{\label{f3:sub2}
    \includegraphics[width=0.46\textwidth]{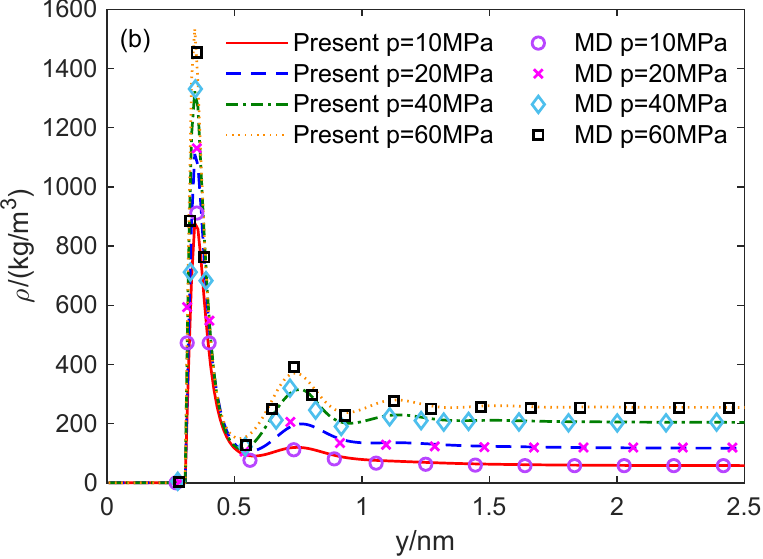}}
    
    \subfloat{\label{f3:sub3}
    \includegraphics[width=0.46\textwidth]{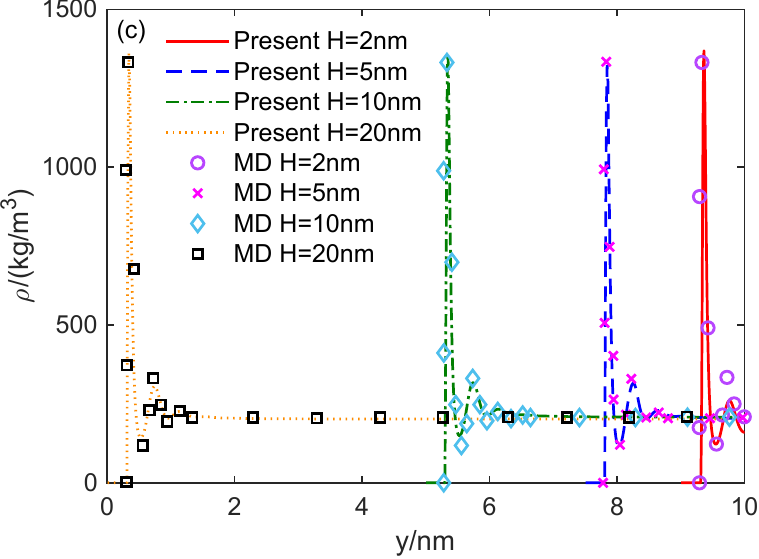}}

  \caption{Density profiles of LJ fluids obtained from the present DUGKS. Effects of $(a)$ temperature $(5\text{nm},\ 40\text{MPa})$; $(b)$ pressure $(5\text{nm},\ 350\text{K})$; $(c)$ channel width $(350\text{K},\ 40\text{MPa})$ on the density distribution. The MD results are given in Ref. \cite{nan2020slip}. Only half domain is shown due to the symmetry.}
  \label{FIG3}
\end{figure}

\subsubsection{Density inhomogeneity}
To assess the accuracy of the present DUGKS, we now consider methane confined between two parallel graphite layers in the $x-z$ plane with different pore sizes $(H)$, pressures $(p)$, and temperatures $(T)$. The spatial domain is discretized using a mesh with $N_{\sigma y}=100$ along the $y$-direction.

Comparisons of density profiles for various values of $H, T,$ and $p$ between numerical results from the proposed DUGKS and the published MD simulations are shown in~\cref{FIG3}, and good agreement can be seen. \cref{FIG3}(a) shows the density profiles at $350$K and $40$MPa in various nanopores. Note that the methane molecules accumulate near the solid wall and form high-density adsorption layers due to the intermolecular interactions. The oscillation amplitude and peak value of the adsorption layer, which are related to the intermolecular interactions, remain nearly constant across different pore sizes. The interesting results imply that the pore size has a negligible effect on the density inhomogeneity under the simulated conditions. \cref{FIG3}(b) presents the density profiles at $40$MPa for various temperatures in $5$nm pore. As temperature increases, the density in the bulk region decreases slightly while the density in the adsorption layer drops more significantly, leading to reduced fluid inhomogeneity. This occurs because, for a constant pressure and volume system, the average number density decreases with temperature according to the equation of state of real gases. Moreover, as the temperature increases, the kinetic energy of fluid molecules increases, enabling more fluid molecules to escape from the absorption layer into the bulk region. We present density profiles at $350$K under various pressures in the $5$nm pore in~\cref{FIG3}(c). It can be seen that densities in both the adsorption layer and the bulk region increase as pressure increases. 

\begin{figure}[H]
  \centering

      \subfloat{\label{f4:sub1}
    \includegraphics[width=0.46\textwidth]{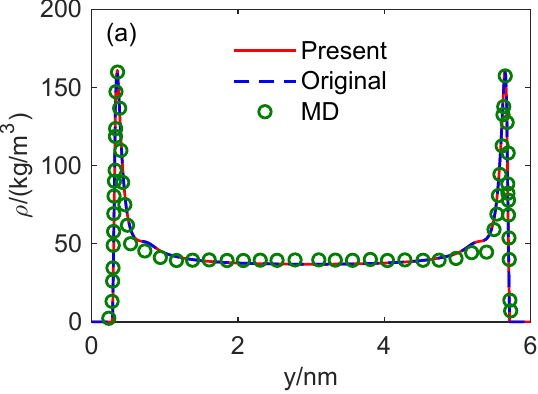}}
    \hspace{0.02\textwidth}
    \subfloat{\label{f4:sub2}
    \includegraphics[width=0.46\textwidth]{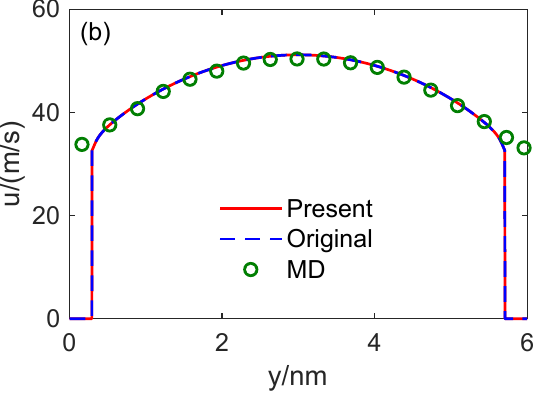}}
  
  \caption{Density $(a)$ and velocity $(b)$ profiles of force-driven flow in organic nanopores obtained from the present DUGKS and the original DUGKS. The MD results can be found in Ref. \cite{chen2017channel}.}
  \label{FIG4}
\end{figure}

\begin{figure}[H]
  \centering

        \subfloat{\label{f5:sub1}
    \includegraphics[width=0.46\textwidth]{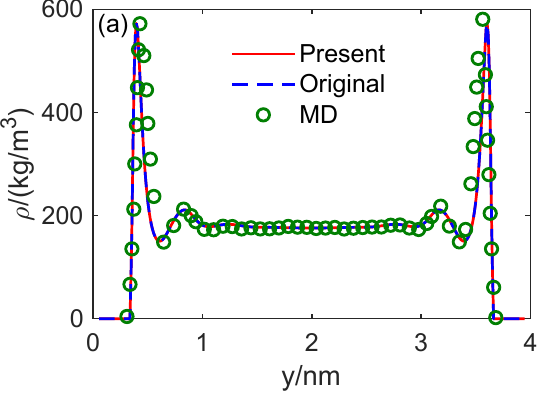}}
    \hspace{0.02\textwidth}
    \subfloat{\label{f5:sub2}
    \includegraphics[width=0.46\textwidth]{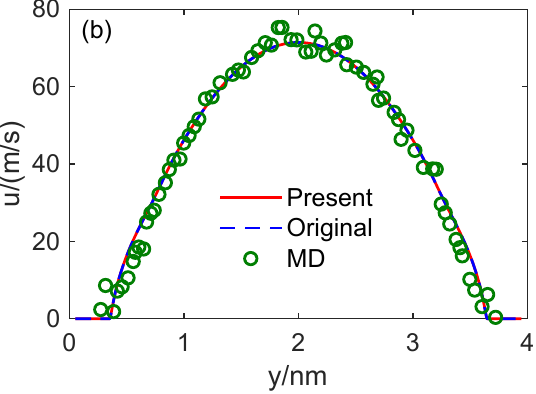}}
  
  \caption{Density $(a)$ and velocity $(b)$ profiles of force-driven flow in inorganic nanopores obtained from the present DUGKS and the original DUGKS. The results of MD simulations are given in Ref. \cite{wang2016breakdown}.}
  \label{FIG5}
\end{figure}

\begin{figure}[H]
  \centering

    \subfloat{\label{f6:sub1}
    \includegraphics[width=0.46\textwidth]{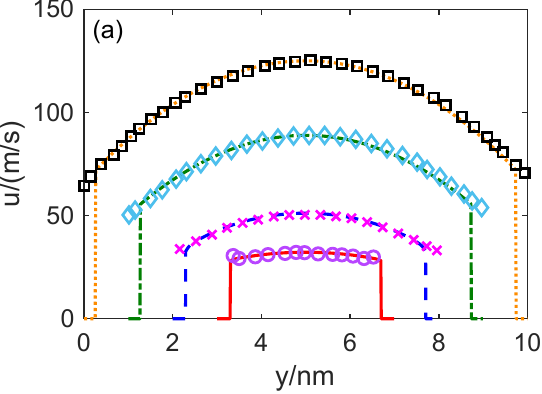}}
    \hspace{0.02\textwidth}
    \subfloat{\label{f6:sub2}
    \includegraphics[width=0.46\textwidth]{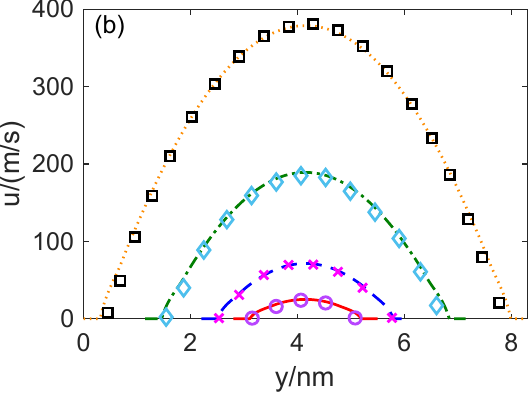}}
    
  \caption{Velocity profiles of force-driven flow. In organic pore $(a)$, from bottom to top, pore size of each line is $4$nm, $6$nm, $8$nm, and $10$nm; in inorganic pore $(b)$, pore size of each line is $2.7$nm, $3.9$nm, $6.0$nm, $8.3$nm. Lines: numerical results from proposed DUGKS; Symbols: MD results taken from Ref. \cite{chen2017channel,wang2016breakdown}.}
  \label{FIG6}
\end{figure}

\subsection{Force-driven flow between two flat plates}
We now consider the force-driven flow of methane in organic kerogen and inorganic matter. The organic and inorganic nanosplits are represented by graphene and calcite, respectively. The entrance and exit effects are pronounced in the pressure-driven MD simulation \cite{suk2017modeling}. To circumvent these effects, a constant external force $F_i$ is applied to each methane molecule to mimic the pressure gradient in MD simulations \cite{chen2017channel,wang2016breakdown}. The relationship between $\bm{F}_i$ and the pressure gradient is given as follows
\begin{equation}
\bm{F}_i=\frac{\nabla p}{n_0}.
\label{eq:46}
\end{equation}
We also employ this force–driven setup to evaluate the accuracy of the proposed DUGKS. The spatial domain is discretized into a mesh with $N_{\sigma y}=200$ along the $y$-direction. The channel length in the $x$ direction $L$ is set to $6\sigma$ and the spatial region is divided by $N_{\sigma x}=20$ along the $x$-direction. A periodic boundary condition is imposed in the $x$-direction. 

The cross-sectional density and velocity profiles in organic kerogen and inorganic pores are presented in~\cref{FIG4} and~\cref{FIG5}, respectively. It can be seen that molecules accumulate around the solid wall. A significant slip velocity at the fluid-solid interface is observed in organic pores, while there is no slip in inorganic pores. This discrepancy can be partly justified based on the difference of wall surface roughness. The surface of the organic pore is smooth, while there is a non-negligible molecular-scale roughness at inorganic pore surface. In organic kerogan pores, the relative $\mathbb{L}_2$ errors between the proposed DUGKS and the original DUGKS for velocity and density are $2.15 \times 10^{-3}$ and $9.21 \times 10^{-3}$, respectively. In inorganic pores, the corresponding errors are $1.87 \times 10^{-3}$ and $8.96 \times 10^{-3}$, respectively. The speedup $Sp$, defined as the ratio of the runtime for $1000$ iteration steps using the original DUGKS to that using the present DUGKS on a single Intel Xeon Gold 6348 2.60GHz CPU, averages 103.51 across the two test cases. The velocity profiles in organic and inorganic pores with different pore sizes are shown in~\cref{FIG6}. The velocity increases with pore size, indicating a stronger mass transport capacity in larger pores. The above results obtained from the proposed DUGKS are in good agreement with the reference results from the MD simulations \cite{chen2017channel,wang2016breakdown}, which shows that the proposed DUGKS can obtain satisfactory predictions. 

\begin{figure}[H]
     \centering
     \includegraphics[width=0.5\textwidth]{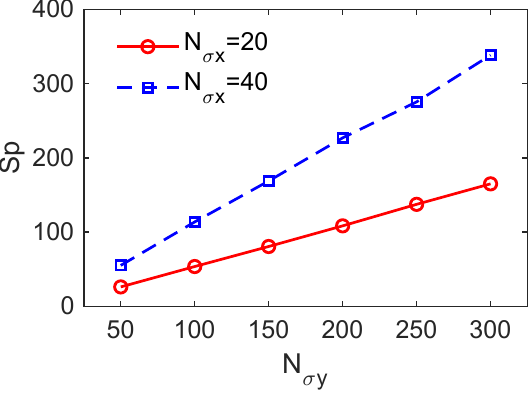}
     \caption{Speedup in force-driven flow with varying $N_\sigma$, compared to the original DUGKS.}
     \label{FIG7}
\end{figure}

\subsection{Comparison of computational efficiency}
The computational costs of the proposed DUGKS and the original one are compared for force-driven flows. The spatial domain is discretized with $N_{\sigma x}=\{20, 40\}$ and $N_{\sigma y}=\{50, 100, 150, 200\}$, which are necessary to ensure the grid independence of the results due to variations in fluid properties along both the $x$- and $y$-directions. The speedup is defined as
\begin{equation}
Sp = \frac{T_o}{T_p},
\label{eq:47}
\end{equation}
where \(T_o\) and \(T_p\) denote the runtimes for \(1000\) iteration steps using the original DUGKS and the present DUGKS, respectively. The corresponding values of \(Sp\) are shown in~\cref{FIG7} as a function of \(N_\sigma = N_{\sigma x} N_{\sigma y} N_{\sigma z}\), with \(N_{\sigma z} = 1\). As expected, the speedup is proportional to the $N_\sigma$. For a fixed $N_{\sigma y}$, the speedup at $N_{\sigma x}=40$ is twice that at $N_{\sigma x}=20$. When $N_{\sigma x}=40$, the speedup can reach up to two orders of magnitude for $N_{\sigma y} > 100$, whereas for $N_{\sigma x}=20$, $N_{\sigma y}$ needs to be increased to $200$ to achieve the same level of speedup. For three-dimensional flows with \(N_{\sigma z} > 1\), the speedup is expected to increase further to three orders of magnitude.

\section{Two-dimensional flows}\label{sec:6}
Based on the excellent performance of the proposed DUGKS for the strongly inhomogeneous kinetic model, we now investigate the two-dimensional flow of methane in inorganic calcite media. The weighting functions for average density are reduced to two-dimensional forms via integration. 

\subsection{Pressure-driven flow between two flat plates}
We first consider pressure-driven flow in a channel of width \(H=10.26\sigma\) and length \(L=5H\). The channel connects two reservoirs maintained at different pressures. The average number density in the outlet reservoir is fixed at \(n_{out}=0.10\,\sigma^{-3}\), while the number density in the inlet reservoir \(n_{in}\) is varied to control the number density ratio \(n_{in}/n_{out}\) and investigate its influence on the flow. The physical domain is discretized on a uniform Cartesian mesh with \(N_{\sigma y}=200\) and \(N_{\sigma x}=20\). The velocity space is discretized by the  abscissa of an \(8\times 8\) half-range Gauss-Hermite quadrature.

Macroscopic constant pressure boundary conditions are not applicable at the nanoscale due to the density inhomogeneity. To describe density inhomogeneity at the nanoscale, we introduce an inhomogeneous pressure boundary condition. At the inlet (\(i=0\)) and outlet (\(i=N_x+1\)) ghost cells, the number density is obtained from the interior cells by linear extrapolation
\begin{equation}
\begin{aligned}
&n'(0,j) = 2\,n(1,j) - n(2,j),\\
&n'(N_x+1,j) = 2\,n(N_x,j) - n(N_x-1,j),
\end{aligned}
\label{eq:48}
\end{equation}
for \(j=1,\ldots,N_y\), where \(N_y\) and \(N_x\) denote the total number of interior fluid cells in the \(y\)- and \(x\)-direction, respectively. Then we rescale the ghost cell densities so that their cross-sectional averages equal the reservoir number densities
\begin{equation}
\begin{aligned}
&n(0,j) = \frac{N_y\, n_{in}}{\displaystyle \sum_{j=1}^{N_y} n'(0,j)}\, n'(0,j),\\
&n(N_x+1,j)  = \frac{N_y\, n_{out}}{\displaystyle \sum_{j=1}^{N_y} n'(N_x+1,j)}\, n'(N_x+1,j).
\end{aligned}
\label{eq:49}
\end{equation}
Then, we impose the nanoscale pressure boundary conditions using a non-equilibrium extrapolation scheme \cite{wu2016discrete,zhao2002non} to ensure a smooth transition of macroscopic fields from the interior to the boundaries.
\begin{figure}[H]
  \centering

    \subfloat{\label{f8:sub1}
    \includegraphics[width=0.46\textwidth]{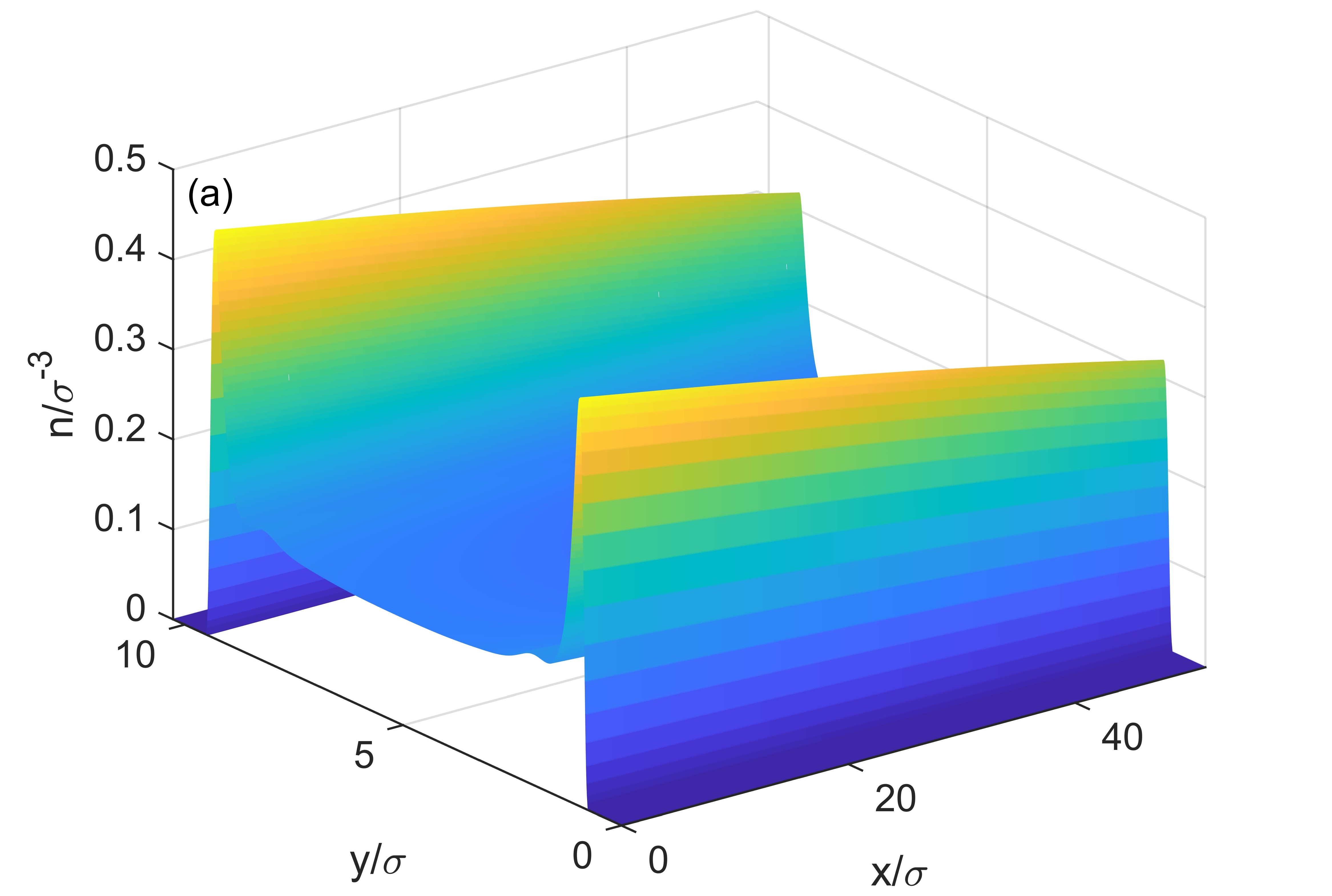}}
    \subfloat{\label{f8:sub2}
    \includegraphics[width=0.46\textwidth]{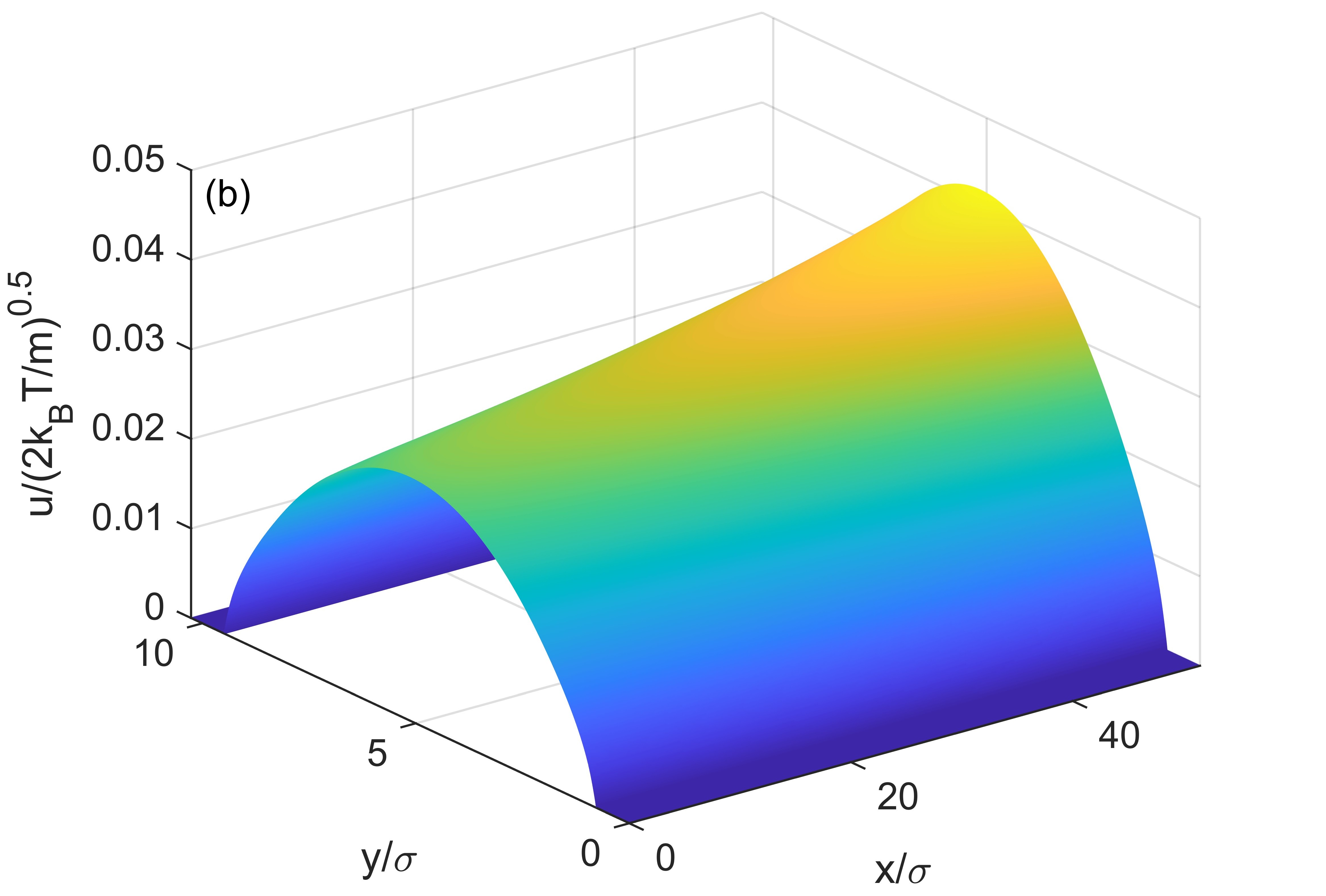}}
    
    \subfloat{\label{f8:sub3}
    \includegraphics[width=0.46\textwidth]{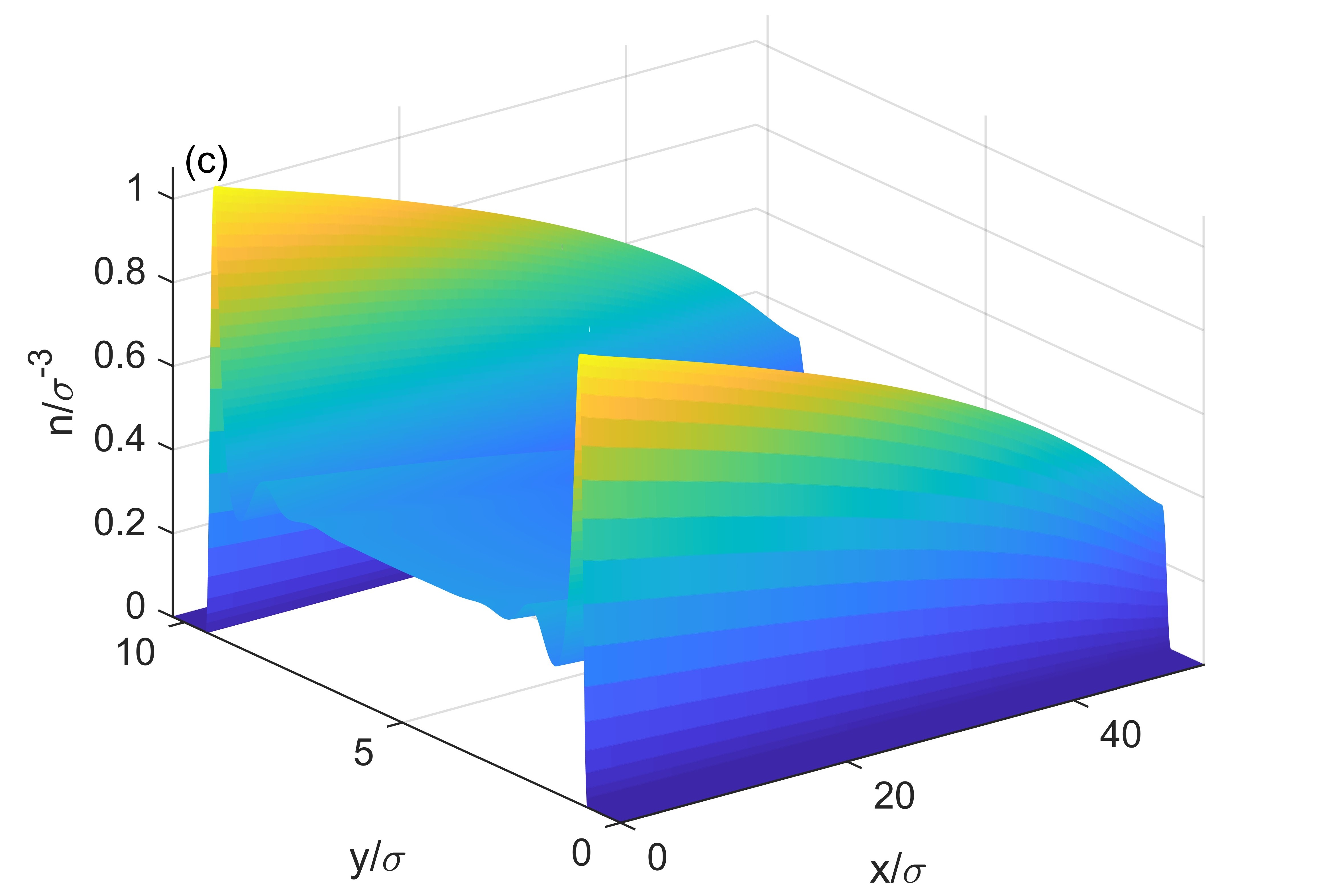}}
    \subfloat{\label{f8:sub4}
    \includegraphics[width=0.46\textwidth]{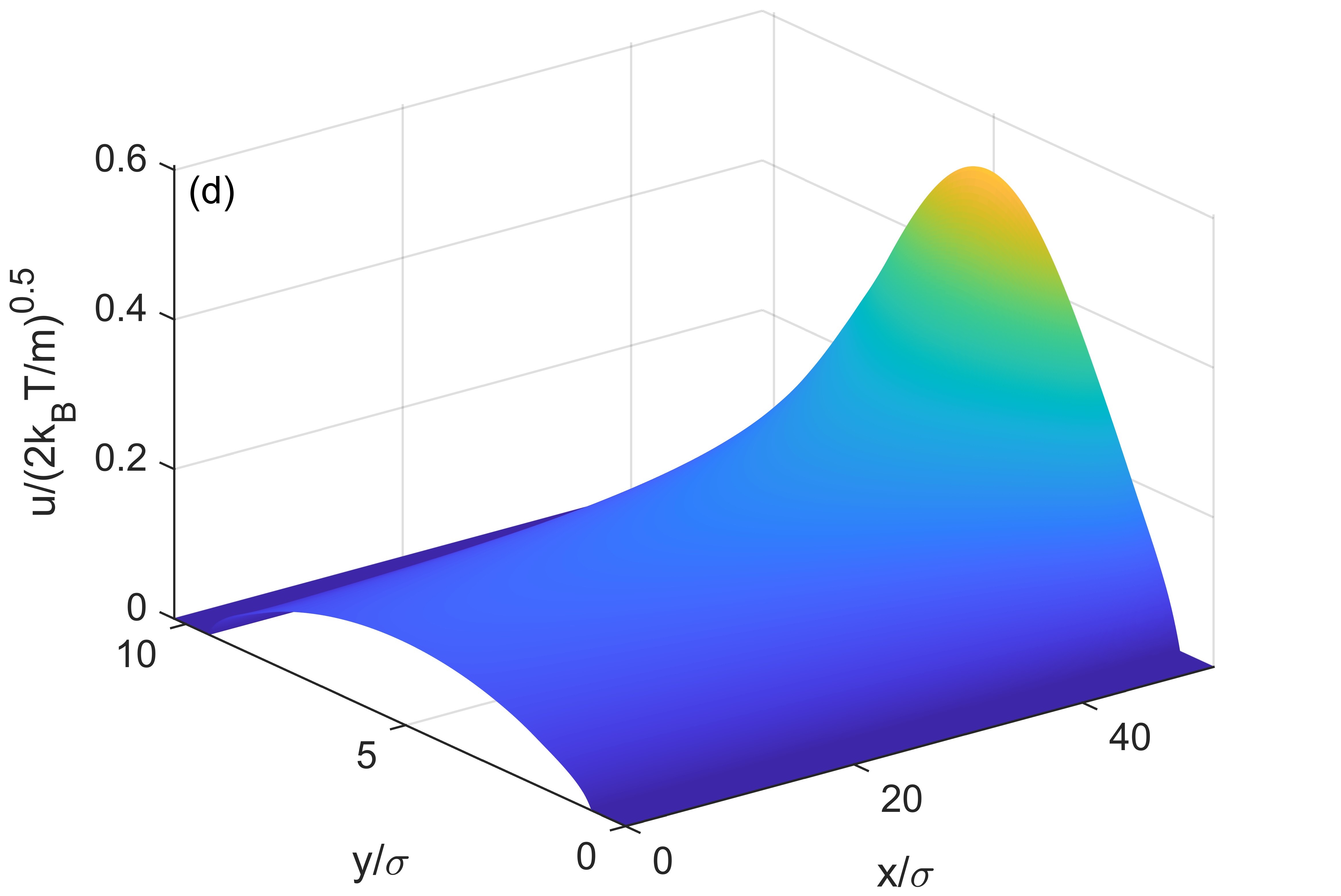}}
    
  \caption{Density (left column) and streamwise velocity (right column) profiles in pressure-driven flow. Top row: \(n_{out}=0.15\,\sigma^{-3}\); bottom row: \(n_{out}=0.40\,\sigma^{-3}\).}
  \label{FIG8}
\end{figure}

\begin{figure}[H]
  \centering
    \subfloat{\label{f9:sub1}
    \includegraphics[width=0.46\textwidth]{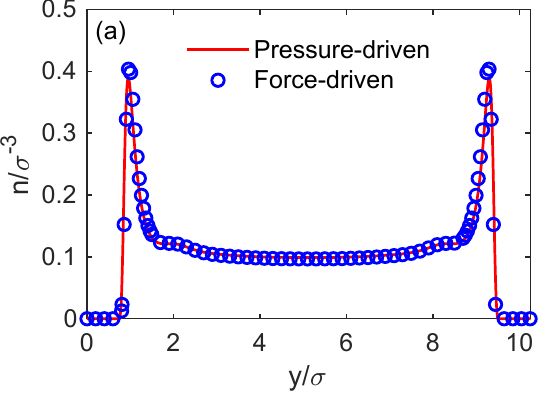}}
    \subfloat{\label{f9:sub2}
    \includegraphics[width=0.46\textwidth]{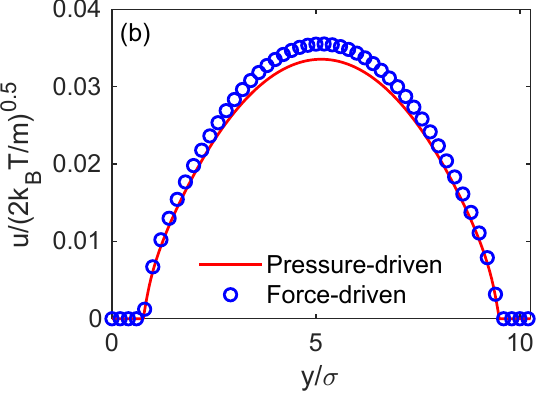}}
    
    \subfloat{\label{f9:sub3}
    \includegraphics[width=0.46\textwidth]{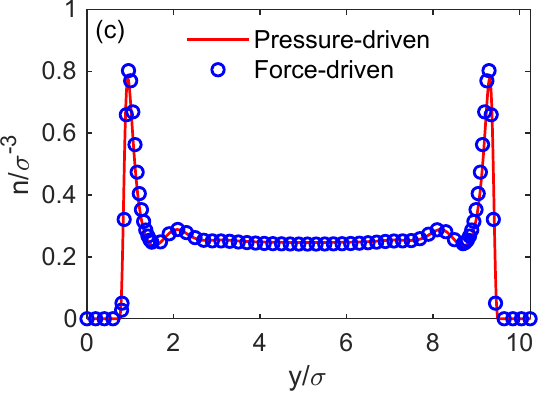}}
    \subfloat{\label{f9:sub4}
    \includegraphics[width=0.46\textwidth]{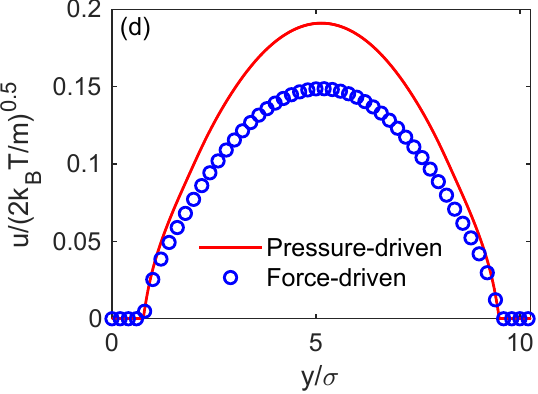}}
  \caption{Pressure-driven vs force-driven flows: streamwise-averaged number density (left column) and velocity (right column). Top row: \(n_{out}=0.15\,\sigma^{-3}\); bottom row: \(n_{out}=0.40\,\sigma^{-3}\).}
  \label{FIG9}
\end{figure}

\begin{figure}[H]
     \centering
     \includegraphics[width=0.5\textwidth]{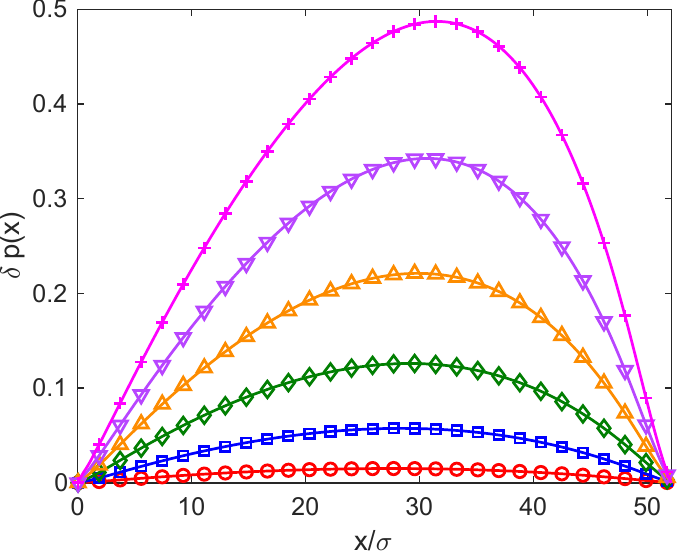}
     \caption{Density deviation \(\delta n(x)\) along the flow direction for \(n_{in}/n_{out}=1.5,\,2.0,\,2.5,\,3.0,\,3.5,\) and \(4.0\). From bottom to top, the curves follow the same order.}
     \label{FIG10}
\end{figure}

\begin{figure}[H]
     \centering
     \includegraphics[width=0.5\textwidth]{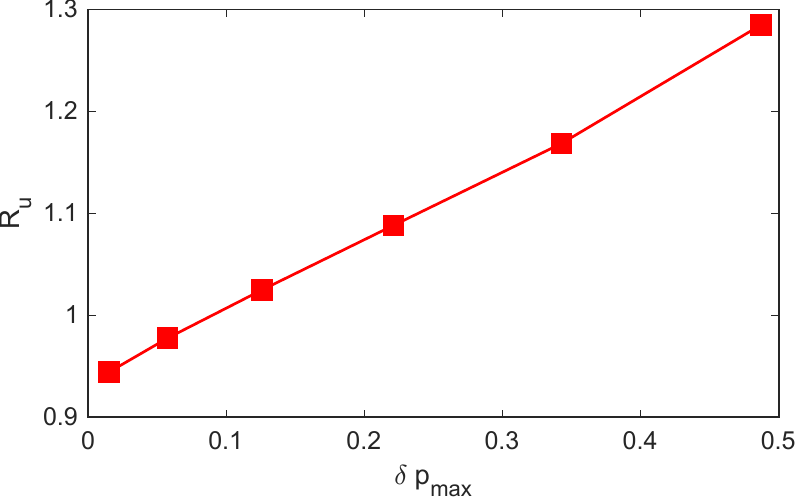}
     \caption{Streamwise-averaged velocity ratio \(R_u\) versus maximum density deviation \(\delta n_{\max}\) across density ratios \(n_{in}/n_{out}\in\{1.5,\,2.0,\,2.5,\,3.0,\,3.5,\,4.0\}\).}
     \label{FIG11}
\end{figure}

\cref{FIG8} shows the density and streamwise velocity profiles for the cases of \(n_{out}=0.15\,\sigma^{-3}\) and \(0.40\,\sigma^{-3}\). The density and streamwise velocity are approximately linear along the flow direction at \(n_{out}=0.15\,\sigma^{-3}\), whereas they become strongly nonlinear at \(n_{out}=0.40\,\sigma^{-3}\). \cref{FIG9} compares the pressure-driven flow with the corresponding force-driven case, with the external force evaluated from Eq.~\eqref{eq:46}. While the density agrees well, the velocity shows noticeable discrepancies owing to a non-negligible deviation from a linear streamwise density distribution at the nanoscale. This conflicts with the assumption in Eq.~\eqref{eq:46} of a uniform streamwise density gradient, i.e., a linear density profile. To quantify the deviation from this linear profile, we define
\begin{equation}
\delta n(x)=\frac{n(x)-n_l(x)}{n_{out}}, 
\quad
n_l(x) = n_{in} + \frac{x}{L}(n_{out}-n_{in}\bigr),
\label{eq:50}
\end{equation}
where \(n(x)\) denotes the cross-sectional average number density and \(n_l(x)\) is the reference linear profile connecting \(n_{in}\) at \(x=0\) and \(n_{out}\) at \(x=L\). \cref{FIG10} shows that the density deviation \(\delta n\) increases with the density ratio \(n_{in}/n_{out}\). We further observe that, at the nanoscale, the streamwise-averaged velocity ratio \(R_u\) between pressure-driven and force-driven cases varies approximately linearly with the maximum number density deviation \(\delta n_{\max}=\max_x \delta n(x)\) over a range of density ratios, as shown in Fig.~\ref{FIG11}. These results demonstrate that pressure-driven and force-driven flows are not generally equivalent at the nanoscale \cite{guan2022evaluation}. 

\subsection{Force-driven flow in a square duct}
We now consider fully developed force-driven flow in infinitely long straight square ducts with widths \(H=10.26\sigma\) and \(H=5.13\sigma\). The duct axis is aligned with the \(z\)-direction and the cross-section lies in the \(x\)-\(y\) plane. Owing to the fully developed assumption, all macroscopic quantities are uniform along the \(z\)-direction, so the problem is effectively two-dimensional in physical space and the simulation is performed on the \(x\)-\(y\) cross-section. In both cases, the cross-sectional average number density is fixed at \(n_{0}=0.30\,\sigma^{-3}\). A constant external force \(\boldsymbol{F}_i=(0,0,F_z)\) with \(F_z=0.032\,\epsilon_{ff}/\sigma\) is applied to each molecule along the \(z\)-direction to drive the flow. The physical domain is discretized by a uniform Cartesian mesh with \(N_{\sigma x}=N_{\sigma y}=200\) cells. Although the spatial domain is treated as two-dimensional, the velocity space remains three-dimensional and is discretized using an \(8\times 8\times 8\) half-range Gauss–Hermite quadrature.

\begin{figure}[H]
  \centering

    \subfloat{\label{f12:sub1}
    \includegraphics[width=0.46\textwidth]{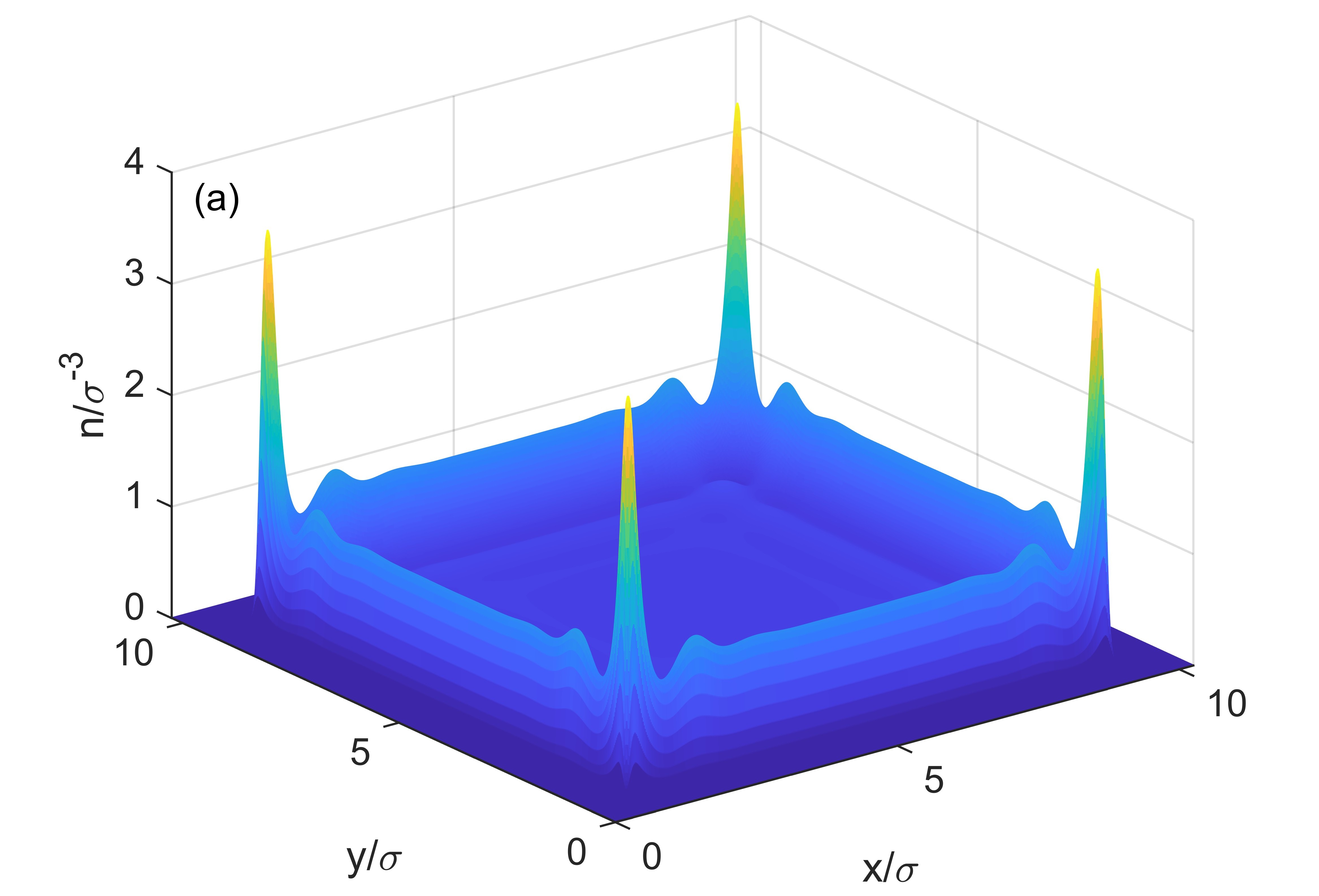}}
    \subfloat{\label{f12:sub2}
    \includegraphics[width=0.46\textwidth]{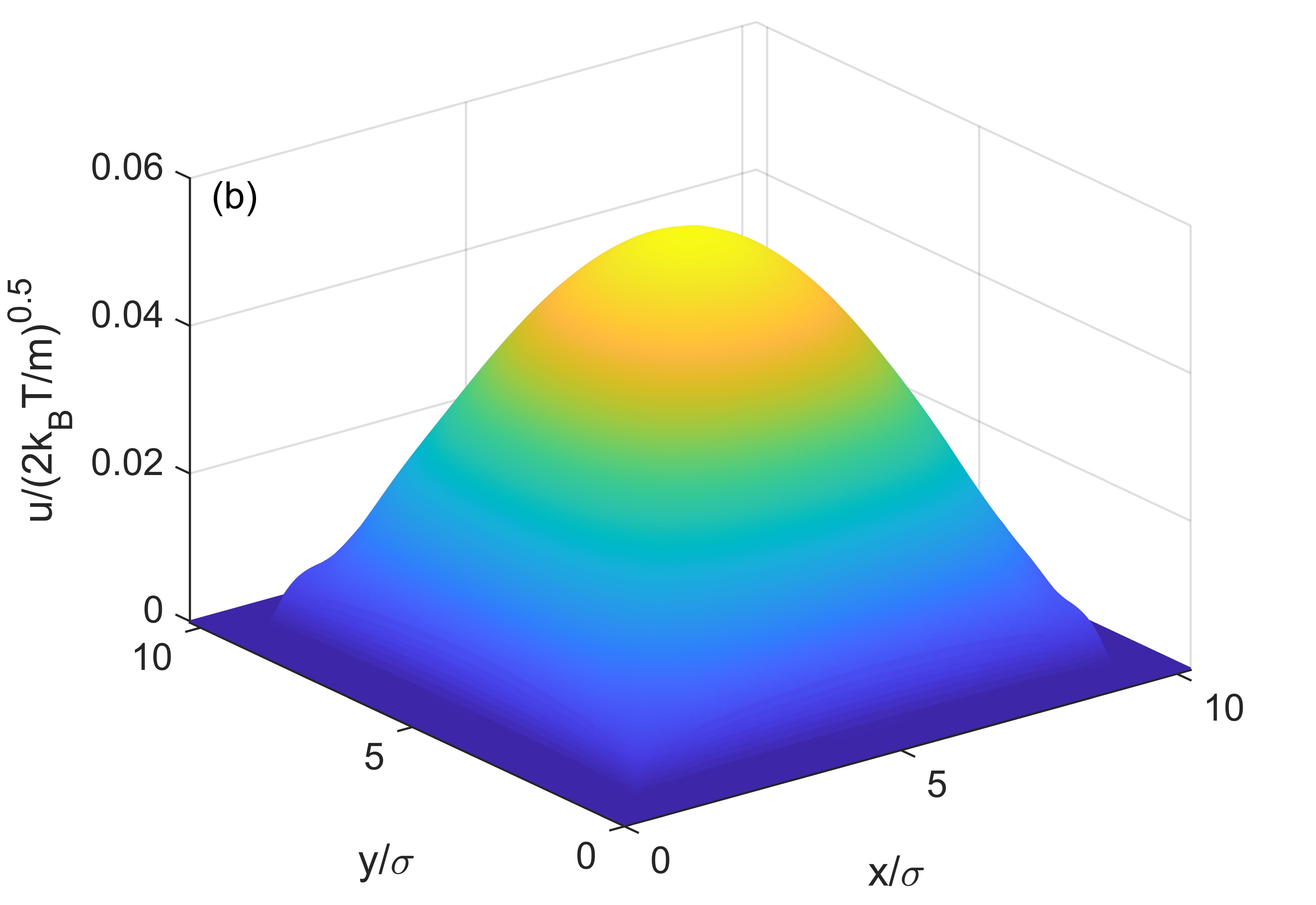}}
    
    \subfloat{\label{f12:sub3}
    \includegraphics[width=0.46\textwidth]{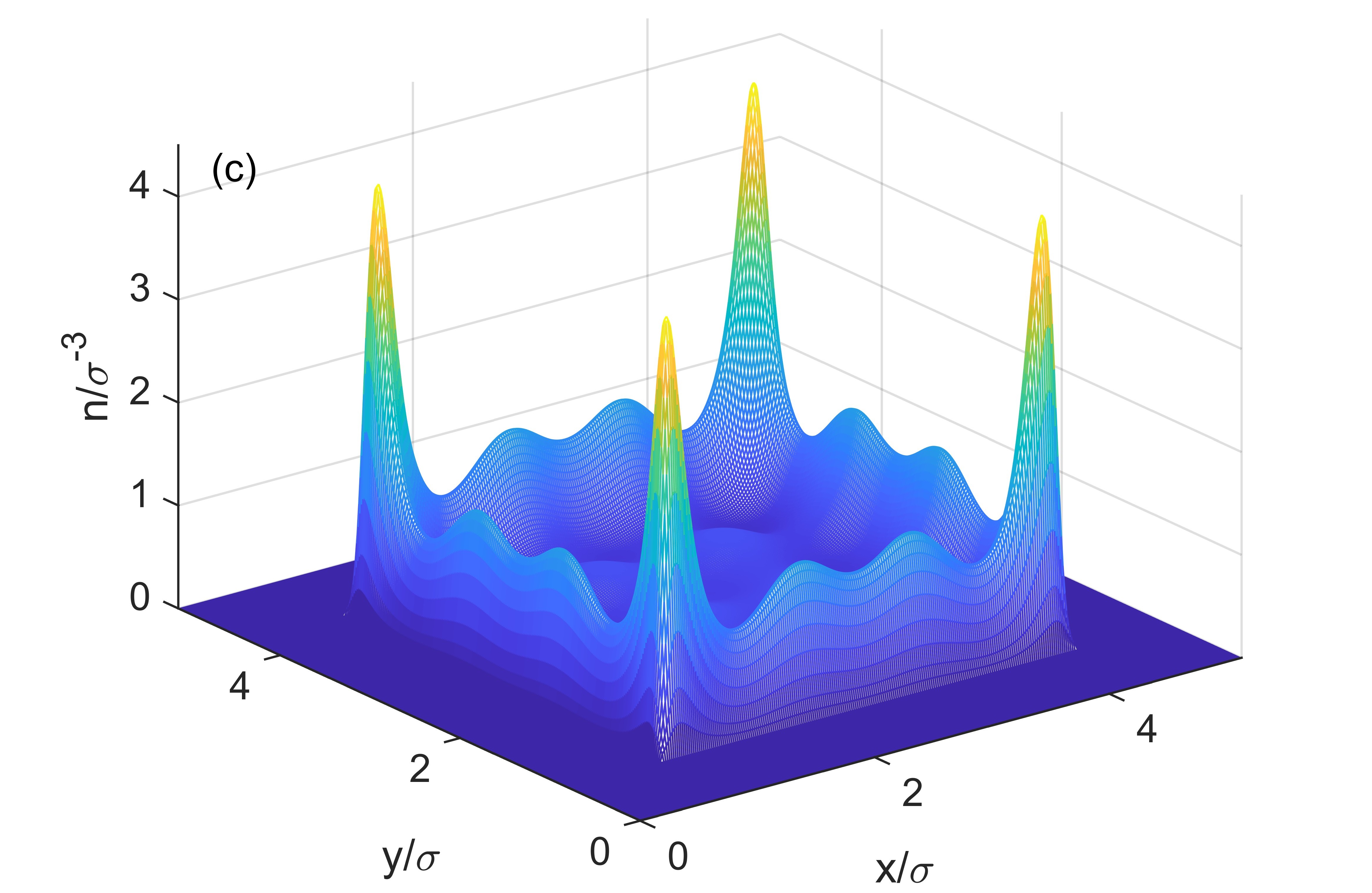}}
    \subfloat{\label{f12:sub4}
    \includegraphics[width=0.46\textwidth]{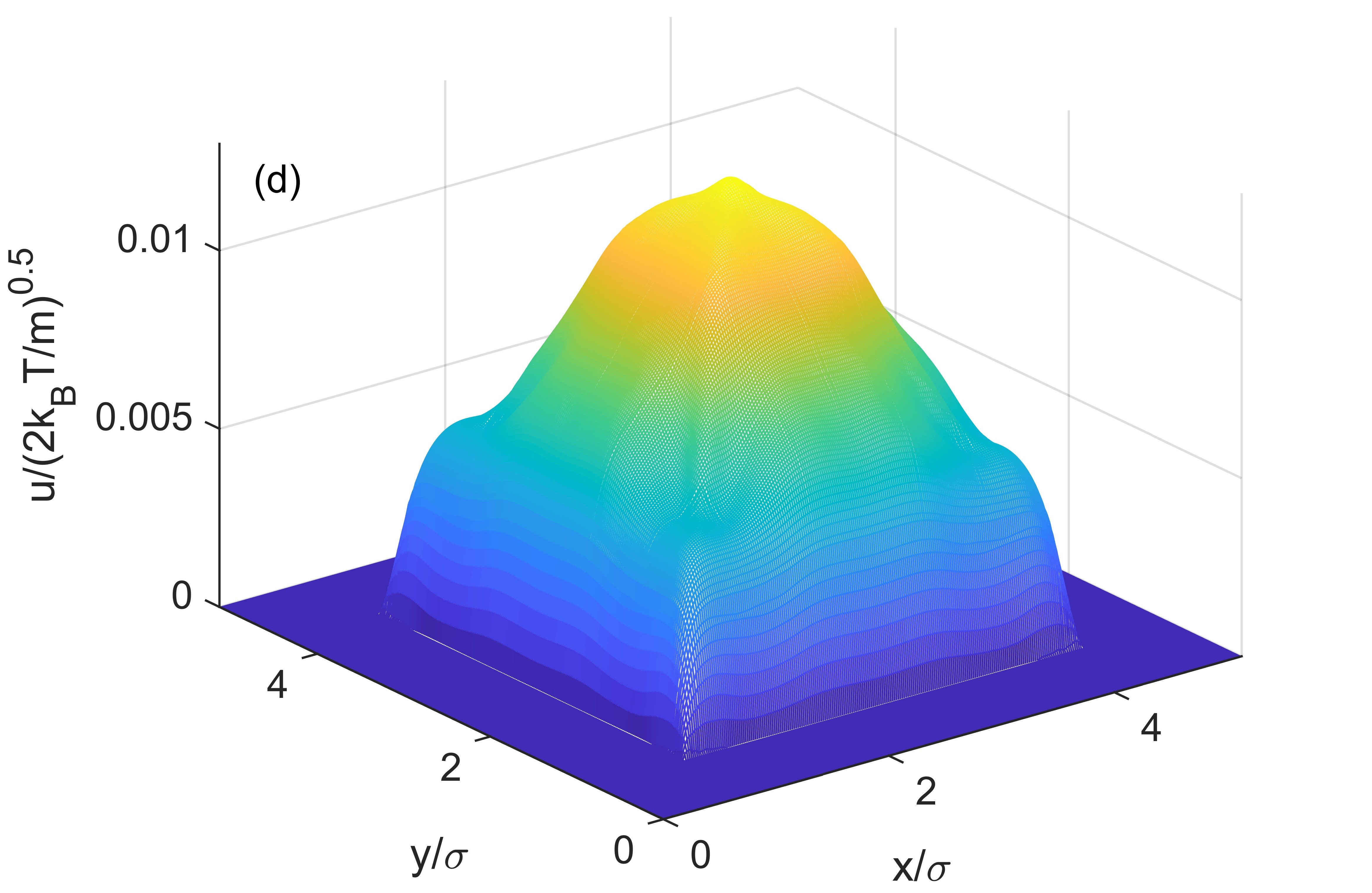}}
    
  \caption{Density (left column) and velocity (right column) profiles for force-driven flow in square ducts. Top row: \(H=10.26\sigma\); bottom row: \(H=5.13\sigma\).}
  \label{FIG12}
\end{figure}

\begin{figure}[H]
  \centering

    \subfloat{\label{f13:sub1}
    \includegraphics[width=0.46\textwidth]{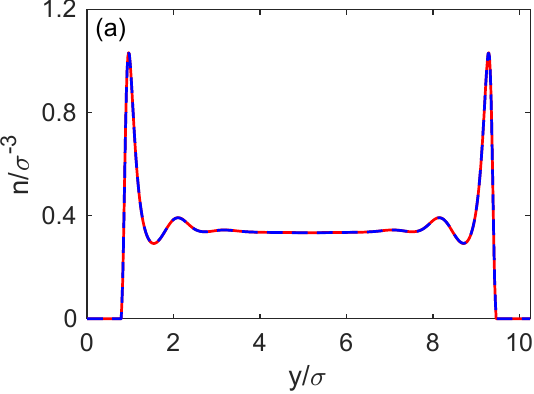}}
    \hspace{0.02\textwidth}
    \subfloat{\label{f13:sub2}
    \includegraphics[width=0.46\textwidth]{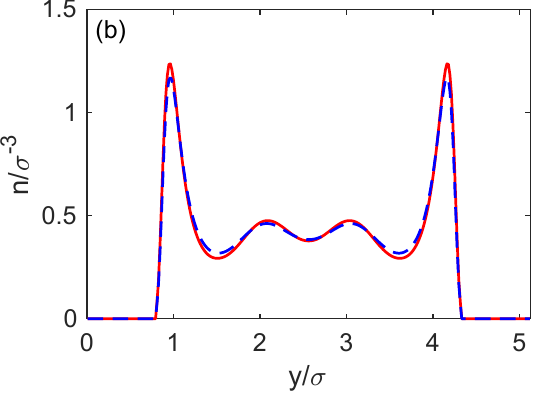}}
    
  \caption{Comparison of the density profile along the centerline of the square duct (red solid) with that in a slit channel (blue dashed) for \(H=10.26\sigma\) and \(H=5.13\sigma\).}
  \label{FIG13}
\end{figure}

\begin{figure}[H]
     \centering
     \includegraphics[width=0.5\textwidth]{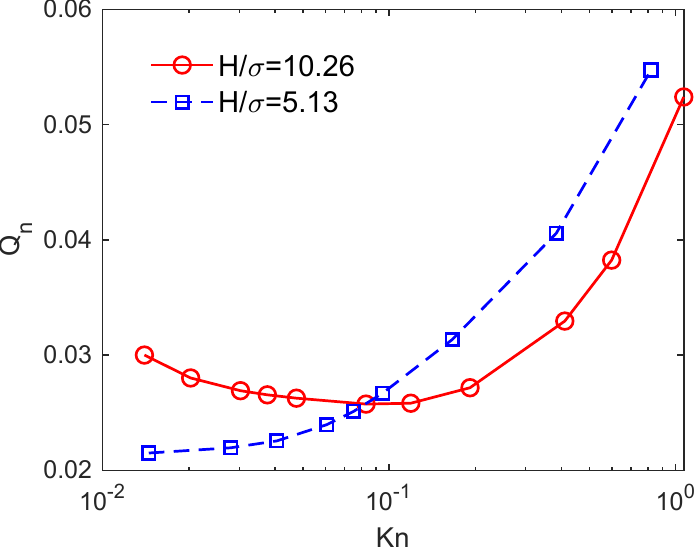}
     \caption{Dimensionless mass flow rate versus Knudsen number in a square duct.}
     \label{FIG14}
\end{figure}

\cref{FIG12} presents the density and streamwise velocity fields in square ducts of different widths. High-density adsorption layers are observed near the solid walls, consistent with our previous results. In particular, molecules accumulate near the corners and form denser adsorption regions due to the combined attraction of the two adjacent walls. In the duct with \(H=10.26\sigma\), the velocity profile is nearly parabolic. By contrast, for \(H=5.13\sigma\) it departs from parabolic behavior owing to stronger density inhomogeneity. \cref{FIG13} compares the density profile along the centerline of a square duct with that in a slit channel of the same width \(H\) and mean number density \(n_0\), and good agreement is observed. The dimensionless mass flow rate \(M_n=M/M_c\) is shown in~\cref{FIG14} as a function of the Knudsen number \(Kn\) for \(H=10.26\sigma\) and \(H=5.13\sigma\). The Knudsen number is defined as the ratio of the molecular mean free path to the duct width
\begin{equation}
Kn=\frac{\lambda}{H}, \quad 
\lambda=\frac{1}{\sqrt{2}\,\pi\, n_0 \sigma^2 \chi(n_0)}.
\label{eq:51}
\end{equation}
The mass flow rate and the reference value are
\begin{equation}
M =  \int_{A} nmu_z dA,
\quad
M_c = n_0mv_cH^2,
\label{eq:52}
\end{equation}
where \(A\) is the cross-sectional area. The characteristic velocity \(v_c\) is taken as
\begin{equation}
v_c=\frac{aH^2}{\nu},\quad a=\frac{F_z}{m},\quad \nu=\frac{\mu(n_0)}{n_0m}
\label{eq:53}
\end{equation}
with \(\nu\) the kinematic viscosity. For \(H=10.26\sigma\), the Knudsen minimum is observed at \(Kn \approx 0.1\); for \(H=5.13\sigma\), the Knudsen minimum disappears and the mass flow rate \(Q_n\) increases monotonically with \(Kn\). It is interesting that the disappearance of the Knudsen minimum has also been observed in previous studies \cite{guo2006generalized}.

\section{Conclusions}\label{sec:7}
The DUGKS developed in Ref. \cite{shan2020discrete} is capable of capturing the flow behaviors of nano-confined strongly inhomogeneous fluids. However, direct computation of multiple integral terms associated with nonlocal gradient, average density, and mean-field force leads to a significant computational cost $O(NN_\sigma)$, where $N$ and $N_\sigma$ represent the number of cells in the entire physical space and integral domain. As a result, previous studies have been limited to one-dimensional problems. In this study, we propose a novel DUGKS that incorporates efficient numerical schemes for these integrals, reducing the computational cost to $O(N)$. A new difference method is introduced to replace the integral method for nonlocal gradients, and an adaptive volume-averaging method for average density is proposed to accurately describe two-dimensional flows. The average density and the mean-field potential are computed on a coarse scale, where fewer points are used to compute the mean-field potential as distance increases due to the rapid decay of attractive potential in space.

The accuracy of the proposed DUGKS has been validated through numerical tests in static fluid structures and flow dynamics. The speedup relative to the original DUGKS has been rigorously calculated. It is shown that the speedup is proportional to $N_\sigma$ and can reach up to two orders of magnitude in two-dimensional flows. These results demonstrate that the proposed DUGKS is a powerful numerical tool for simulating nanoscale flows.

We further investigate two-dimensional flows to reveal the exotic phenomena at the nanoscale. In the pressure-driven channel flow, we find that pressure-driven and force-driven flows are generally not equivalent because the density distribution along the streamwise direction is nonlinear. In the force-driven square duct flow, density peaks are observed near the corners due to the combined attraction from adjacent walls. In future work, we plan to apply the proposed DUGKS to complex three-dimensional and thermal flows.

\section*{Declaration of Competing Interest}
No Conflict of Interest declared.
\section*{Acknowledgments}
This work was supported by the National Natural Science Foundation of China (Grant No. 12472290). The computation is performed in the HPC Platform of Huazhong University of Science and Technology.

\appendix
\section{Reduced weighting functions for average densities}
The average densities obtained from each method can be reduced to one- and two-dimensional forms by integration. The reduced average densities $\bar{n}(y)$ and $\bar{n}(x,y)$ are given by
\begin{equation}
\begin{aligned}
& \bar{n}(y)=\int \omega_{y'}(y') n(y+y') \, dy',\\
& \bar{n}(x,y)=\int \omega_{x'y'}(x',y') n(x+x',y+y') \, dx'dy',
\end{aligned}
\end{equation}
where the reduced weighting functions $\omega_{y'}(y')$ and $\omega_{x'y'}(x',y')$ are defined by
\begin{equation}
\begin{aligned}
& \omega_{y'}(y')=\int  \omega(\bm{r}') \, dx'dz',\\
& \omega_{x'y'}(x',y')=\int \omega(\bm{r}') \, dz'.
\end{aligned}
\end{equation}
The reduced weighting functions corresponding to the averaging methods discussed in~\cref{sec:4} are the VAM:
\begin{equation}
\begin{aligned}
& \omega_{y'}(y')=\frac{3}{4\sigma^3} H(\sigma-|y'|) (\sigma^2-y'^2),\\
& \omega_{x'y'}(x',y')=\frac{3}{2\pi\sigma^3} H(\sigma^2-x'^2-y'^2) (\sqrt{\sigma^2-x'^2-y'^2});
\end{aligned}
\end{equation}
AVAM:
\begin{equation}
\begin{aligned}
& \omega_{y'}(y')=\frac{H(\sigma-|y'|) (\sigma^2-y'^2)}{\int H(\sigma-|y'|) (\sigma^2-y'^2) \,dy'},\\
& \omega_{x'y'}(x',y')=\frac{H(\sigma^2-x'^2-y'^2) (\sqrt{\sigma^2-x'^2-y'^2})}{\int H(\sigma^2-x'^2-y'^2) (\sqrt{\sigma^2-x'^2-y'^2}) \,dx'dy'};
\end{aligned}
\end{equation}
Tarazona's method:
\begin{equation}
\omega_{y'}(y')=\omega_{0y'}(y')+\omega_{1y'}(y')\bar{n}(y + y')+\omega_{2y'}(y')\bar{n}(y + y')^2;
\end{equation}
where
\begin{equation}
\begin{aligned}
\omega_{0y'}(y')= & \frac{3}{4 d^3} H(d-|y'|)\left(d^2-y'^2\right) ,\\
\omega_{1y'}(y')= & H(2 d-|y'|) H(|y'|-d) 2 \pi\left[0.288 d(2 d-|y'|)-\left(\frac{0.924}{2}\right)\left(4 d^2-|y'|^2\right)+\left(\frac{0.764}{3 d}\right)\left(8 d^3-|y'|^3\right)\right. \\
& \left.-\left(\frac{0.187}{4 d^2}\right)\left(16 d^4-|y'|^4\right)\right]+H(d-|y'|) 2 \pi\left\{\frac{0.475}{2}\left(d^2-|y'|^2\right)-\frac{0.648}{3 d}\left(d^3-|y'|^3\right)\right. \\
& \left.+\frac{0.113}{4 d^2}\left(d^4-|y'|^4\right)+d^2\left[0.288-\frac{3}{2}(0.924)+\frac{7}{3}(0.764)-\frac{15}{4}(0.187)\right]\right\},\\
\omega_{2y'}(y')= & H(d-|y'|) \frac{10 \pi^2 d^3}{144}\left[3\left(d^2-|y'|^2\right)-\frac{4}{d}\left(d^3-|y'|^3\right)+\frac{5}{4 d^2}\left(d^4-|y'|^4\right)\right] .
\end{aligned}
\end{equation}
The average density in Tarazona's method cannot be reduced to two-dimensional form due to the complexity of the weighting function.

\bibliographystyle{elsarticle-num}
\bibliography{reference}
\end{document}